\documentclass[12pt,preprint]{aastex}

\shorttitle{NUclei of GAlaxies}
\shortauthors{Haan et al.}

\begin{document}
   \title{Atomic Hydrogen Properties of AGN Host Galaxies:\\
 HI in 16 NUclei of GAlaxies (NUGA) Sources}

   \author{Sebastian Haan}
   \affil{Max-Planck-Institut f\"ur Astronomie (MPIA),
              K\"onigstuhl 17, 69117 Heidelberg, Germany}
   \email{haan@mpia.de}

   \author{Eva Schinnerer}
   \affil{Max-Planck-Institut f\"ur Astronomie (MPIA),
              K\"onigstuhl 17, 69117 Heidelberg, Germany}
   \email{schinner@mpia.de}

   \author{Carole G. Mundell}
   \affil{Astrophysics Research Institute, Liverpool John Moores University, Twelve Quays House, Egerton Wharf, Birkenhead, CH41 1LD, UK}
   \email{cgm@astro.livjm.ac.uk}

   \author{Santiago Garc{\'{\i}}a-Burillo}
   \affil{Observatorio Astronomico Nacional (OAN)-Observatorio de Madrid,
	     Alfonso XII, 3, 28014-Madrid, Spain}
   \email{burillo@oan.es}

   \author{Francoise Combes}
   \affil{Observatoire de Paris, DEMIRM, 61 Av. de l Observatoire,
	     75914-Paris, France}
   \email{francoise.combes@obspm.fr}

\begin{abstract}

We present a comprehensive spectroscopic imaging survey of the distribution and kinematics of atomic hydrogen (HI) in 16 nearby spiral galaxies hosting low luminosity AGN, observed with high spectral and spatial resolution (resolution: $\sim$20\arcsec, $\sim$5~km~s$^{-1}$) using the NRAO Very Large Array (VLA). The sample contains a range of nuclear types, ranging from Seyfert to star-forming nuclei and was originally selected for the NUclei of GAlaxies project (NUGA) - a spectrally and spatially resolved interferometric survey of gas dynamics in nearby galaxies designed to identify the fueling mechanisms of AGN and the relation to host galaxy evolution. Here we investigate the relationship between the HI properties of these galaxies, their environment, their stellar distribution and their AGN type. The large-scale HI morphology of each galaxy is classified as ringed, spiral, or centrally concentrated; comparison of the resulting morphological classification with AGN type reveals that ring structures are significantly more common in LINER than in Seyfert host galaxies, suggesting a time evolution of the AGN activity together with the redistribution of the neutral gas. Dynamically disturbed HI disks are also more prevalent in LINER host galaxies than in Seyfert host galaxies. While several galaxies are surrounded by companions (some with associated HI emission), there is no correlation between the presence of companions and the AGN type (Seyfert/LINER).
\end{abstract}


%

\section{Introduction}
\label{sec:intro}
Active Galactic Nuclei (AGN) represent some of the most extreme conditions and include the most powerful individual objects to be found throughout the Universe. Nowadays, the phenomena of nuclear activity is generally understood to be the result of accretion of material onto a SuperMassive Black Hole (SMBH); very likely through the infall of gas from its host galaxy \citep[e.g.][]{Rees}. Although SMBHs exist in most galaxies, only a small fraction of all galaxies exhibit nuclear activity \citep{Huchra,Ho97b, Miller}. The majority of AGNs show low-ionization narrow emission-line regions (LINERs), which have luminosities lower than Seyfert galaxies and quasars. Originally, LINERs and Seyferts were considered to form a continuous distribution in the usual line-ratio classification diagrams. Recent investigations of the host properties of emission-line galaxies from the Sloan Digital Sky Survey \citep{Kew06} confirmed that their nuclear activities are due to accretion by a SMBH. But interestingly, LINERs and Seyferts are also clearly separable in emission line ratio diagrams \citep{Gro06}, and that these two classes have distinct host properties \citep{Gro06}, implying that their distinction is in fact not simply an arbitrary division.
An open question is therefore what mechanisms give rise to the different AGN types observed.
\par
A coherent picture of AGN feeding, e.g. how to remove the angular momentum from host galaxy gas so that it can reach the central parsec, is still missing. Therefore it is expected that a hierarchy of mechanisms might combine to transport gas from large kpc scales down to the inner pc scales \citep{Shl90, Com03, Jog04}.
Additionally, trigger mechanisms driven by the galactic environment, like minor mergers and tidal forces, might play a role as well \citep{Barnes}.
A larger gas fraction for Seyferts than for normal galaxies was suggested by \cite{Hunt99a} and a prevalence of stellar rings (RC3 classification) in active galaxies (Seyfert and LINERs) compared to non-active galaxies was suggested \citep{Hunt99b}. NICMOS imaging of the centers of 250 nearby galaxies \citep{Hunt04} revealed systematic morphological differences between HII/starburst (most disturbed), Seyfert (intermediate disturbed), LINER and normal galaxies (most regular). The study of the circumnuclear regions of 24 Seyfert2 galaxies using HST images \citep{Mar99} revealed a dominance of nuclear spirals, suggesting that spiral dust lanes may be responsible for feeding gas to the central engines. \par

Theoretical simulations have also made progress in addressing the nature of nuclear fueling with respect to different types of gravitational instabilities and their feeding efficiency: nested bars \citep[e.g.][]{Shl89, Friedli, Mac00, Eng04}, gaseous spiral density waves \citep[e.g.][]{Eng00, Mac02, Mac04a, Mac04b}, m = 1 perturbations \citep[e.g.][]{Shu, Jun96, Gar00} and nuclear warps \citep{Sch00} have all been suggested as possible transport mechanisms. \par

In order to distinguish models for nuclear fueling, observations of neutral gas with high angular and velocity resolution are required. Therefore, the IRAM key project NUclei of GAlaxies \citep[ NUGA; see][]{Gar03} was established - a spectroscopic imaging survey of gas in the centers of nearby low luminosity AGN. As most of the gas in galaxy nuclei is in the molecular phase, the survey used millimeter CO
lines to conduct a detailed mapping of molecular gas dynamics at {\em high-resolution} (0.5$\arcsec$) in the central kiloparsec of AGN hosts. 
The CO-NUGA survey (using the IRAM Plateau de Bure Interferometer) reveals a wealth of nuclear gas distribution and kinematics which are studied in detail: (a) m=1 gravitational instabilities (one-arm
spirals and lopsided disks; NGC~4826: \cite{Gar03b}), (b) m=2 instabilities (two-arm spiral wave) expected to form in stellar bar potentials show only small amounts of molecular gas coincident with the AGN (e.g. NGC~4569: \cite{Boo07}, NGC~4579: \cite{Gar05} NGC~6951: Schinnerer et al., in prep.), (c) stochastic spirals that are related to non self-gravitating instabilities, and rings \citep[NGC~7217:][]{Com04}, and (d) large scale warps that might extend into the central kiloparsec \citep[NGC~3718:][]{Kri05}. Besides the CO studies also extended radio continuum components resembling jet emission from AGN were detected \citep{Kri07a} and a molecular gas disk/torus of dense gas was observed in the HCN line emission \citep[NGC~6951:][]{Kri07b}. Additionally, Garc{\'{\i}}a-Burillo et al.
(\citeyear{Gar05}) derived in a pilot study the gas inflow rates via measurements of the gravitational torque onto the molecular gas in the central kiloparsec for 4 galaxies. \par

To complement these CO data and provide a more complete census of the ISM and gas flows from the outskirts of the galaxy disks to the very center, the HI-NUGA project was initiated. HI-NUGA provides observations of the HI gas distributions and kinematics for 16 galaxies (including archival data for 2 galaxies) of the NUGA sample using NRAO's Very Large Array (VLA).
Determining the HI distribution and kinematics is necessary to establish whether characteristics of the nuclear dynamics, and hence the nuclear activity, depend on overall host galaxy properties. Possible dependencies might also exist between nuclear modes and large scale drivers (i.e. tidal forces exerted by companions), or the overall content/distribution of the atomic gas. 
Since the HI gas extends in most disk galaxies much further than the optical disk, it is more loosely bound in the outer region because of the decrease of the gravitational potential.  Hence, it provides an ideal tool to identify interactions and tidal features \citep[e.g.,][]{Sim87, Mun95}.
Furthermore, due to its dissipative nature the gas is more sensitive to dynamical disturbances, both internal ones such as non-axisymmetric potentials or external ones (e.g. tidal interactions). Thus, one could expect, for example, to find a prevalence of Seyfert activity with asymmetries in the gas distribution and kinematics.\par

In this paper we present for 16 nearby active galaxies data obtained in the 21 cm emission line of neutral hydrogen using the VLA. The relation between HI properties, such as the large scale environment (HI/optical companions, HI disk asymmetries) and the AGN type (Seyfert, LINER, HII) are analyzed in detail. The outline of the paper is as follows: We describe the HI-NUGA sample, the observation and data reduction in \S \ref{sec:obs}. The results and derived HI properties are presented in \S \ref{sec:res}, along with a discussion of the presence of companions and tidal disturbances of the HI disks and any correlation with AGN type. A comparison of AGN types and optical (stellar) light distribution is also examined. The results are discussed in the context of correlations with the HI environment (e.g. tidal forces) and the AGN type (\S \ref{sec:dis}). A summary is presented in \S \ref{sec:sum}.

\section{Sample Description, Observations and Data Reduction}
\label{sec:obs}

\subsection{Sample Selection and Survey Goals}
\label{subsec:obs_sample}
The NUGA project is a multiwavelength study of the nuclei of nearby AGN. The core sample \citep{Gar03} consists of 12 nearby AGN with D $\leq$ 54Mpc selected for the availability of high-quality optical and near infrared (NIR) images obtained with both ground-based telescopes and the Hubble Space Telescope (HST). The distribution and kinematics of molecular gas in all galaxies in the sample have been observed with the IRAM PdBI at 1mm and 3mm with maximum angular and spectral resolutions of ~0.5$\arcsec$ and 3.6 km s$^{-1}$, respectively. The NUGA supersample adds another 16 galaxies with CO data of similar quality obtained by the NUGA team with the IRAM array as part of other projects.  
\par
For our detailed HI study using the VLA, 16 galaxies were selected (15 of which have high quality CO data of $\leq1''$ resolution): the entire CO-NUGA parent sample, 3 additional galaxies from the CO-NUGA supersample and NGC 5248. The HI-NUGA sample therefore covers a variety of nuclear activity: 7 Seyferts, 7 LINERs and 2 HII galaxies. 
All galaxies in our sample are spiral galaxies, ranging in Hubble type from Sa to Sbc in disk type. Typical spatial resolution of the HI data is 21$\arcsec$=(1.9 kpc) for a mean distance of $\sim$20 Mpc; the distance of our sample ranges from 4 Mpc to 54 Mpc. 
An overview of the general properties of our HI sample is given in Tab.~\ref{table_intro}. 
The overall aims of the HI-NUGA project are to 1) determine the global HI properties of the sample galaxies and their large scale environment to search for relations between the nuclear activity and the HI environment, for all disk properties of the full sample, 2) to trace the gas inflow on various spatial scales for the best candidates of the HI-NUGA sample with high angular resolution HI data (VLA B+CD array) and to determine the fueling rates by using gravity torques,
3) to obtain detailed dynamical models (using SPH codes with a fixed gravitational potential and/or N-body codes.) describing the gas flow in the galactic disk over several radii for the most representative galaxies. 
In this paper we concentrate on goal 1).

\subsection{AGN classification}
\label{subsec:obs_agnclass}
Active galaxies can be separated into the categories of emission-line AGNs and HII region–like galaxies (starburst) based on the differing nature of the photoionization source. 
Emission-line AGNs can be divided into Seyferts and LINERs. The majority of AGN are LINERs with nuclear optical spectra that are dominated by emission lines of low-ionization species such as [OI] $\lambda$6300, [OII] $\lambda$3726,9 and [SII] $\lambda$6717 \citep{Ho97a}. The main difference between Seyferts and LINERs is thought to be accretion rate, which explains why LINERs have lower AGN luminosities than Seyfert galaxies. 
Although Seyferts and LINERs are basically powered by the same engine, i.e. accretion onto a SMBH, the Seyfert/LINER distribution is bimodal \citep{Gro06}, implying different accretion mechanisms \citep[for a general review see][]{Ulr97, Com01, Net07}.
Furthermore, LINERs seem in general to be hosted by older, more massive, less dusty galaxies with higher stellar velocity dispersions and lower nuclear [OIII] luminosities than Seyfert galaxies \citep{Kew06}. Despite numerous studies in this field of research, the origin of their differences (e.g. different stages in evolution) is still under debate. \par

\begin{figure}[ht]
\begin{center}
\resizebox{1\hsize}{!}{\includegraphics[bb=18 159 580 673]{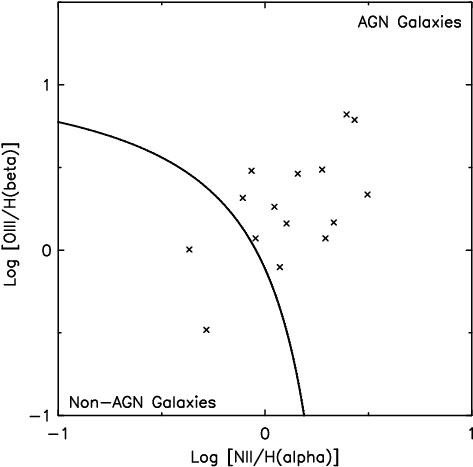}
\includegraphics[bb=18 159 580 673]{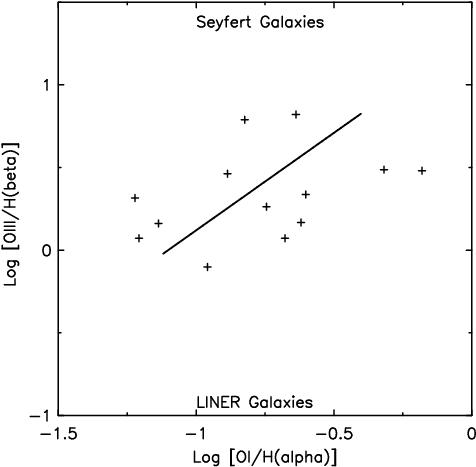}}
\caption{\footnotesize
AGN classification for our sample following the definition of Kewley et al. (\citeyear{Kew06}). The left panel shows the separation between AGN (upper right part) and non-AGN (lower left part) galaxies using the optical line ratios of [OIII]/H$\beta$ versus [NII]/H$\alpha$ as described in the text. In the right panel the dividing line between Seyfert (top part) and LINERs (bottom part) is indicated by using the [OIII]/H$\beta$ versus [OI]/H$\alpha$ line ratios (see text). Note that NGC~6574 is excluded in these diagrams, as no accurate ratios of the lines used were available in the literature. We used the classification as Seyfert given by NED. 
\label{fig_Kew}}
\end{center}
\end{figure}

Although all the galaxies in our sample have several nuclear classifications available in the literature, we reclassified their nuclear types as described in the following, in order to have a coherent scheme for our sample. The most common method to select star-forming galaxies is based on the Baldwin, Phillips \& Terlevich (\citeyear{Baldwin}) empirical diagnostic diagrams using the optical line ratios of [OI]/H$\alpha$, [SII]/H$\alpha$, [NII]/H$\alpha$, and [OIII]/H$\beta$. For the purpose of this paper we separated our sample in Seyferts, LINERs and non-AGN (HII+Transitions) objects following the most recent method of Kewley et al. (\citeyear{Kew06}):  First we distinguished AGN galaxies from non-AGN galaxies by using the optical line ratio [OIII]/H$\beta$ versus [NII]/H$\alpha$ (left panel Fig.~\ref{fig_Kew}). The line ratios are taken from the cataloged data of Ho et al. (\citeyear{Ho97a}) and Kim et al. (\citeyear{Kim95}). For NGC~6574 no accurate line ratios were available in the literature for this method and we used its classification listed in NED as Seyfert. The line separating AGN and Starburst/Transition galaxies is given by \citep{Kew06}:
\begin{equation}
log(\frac{[OIII]}{H\beta})=\frac{0.61}{log(\frac{[NII]}{H\alpha})-0.47}+1.19
\end{equation}
Secondly, we divided the AGN galaxies in Seyfert and LINERs by using the [OIII]/H$\beta$ versus [OI]/H$\alpha$ diagram (right panel Fig.~\ref{fig_Kew}) with a separating line defined by:
\begin{equation}
log(\frac{[OIII]}{H\beta})=1.18 \cdot log(\frac{[OI]}{H\alpha})+1.30 .
\end{equation}   
The classification of our sample results in 2 non-AGN galaxies (HII and transition objects), 7 Seyferts and 7 LINERs. 
Since only 2 non-AGN galaxies are present in our sample, we will mainly focus on a comparison between the properties of Seyferts and LINERs. 
A detailed comparison of HI properties between Seyfert and non-AGN galaxies is carried out by Mundell et al. (in prep.). The average morphology type taken from Hyperleda is roughly the same for Seyferts (t=2.81) and LINERs (t=2.69). Hence, effects due to bias of the morphology type are unlikely.

\subsection{VLA Observations and Data Reduction}
\label{subsec:obs_data}

The 16 galaxies presented here were observed in 2003 and 2004 in their HI line emission using the VLA in its C- and D-configuration. Three galaxies (NGC~3627, NGC~4736, NGC~4826) were also observed as part of the THINGS project (Walter et al., in prep., BC-array data) while adequate data for NGC~2782 and NGC~3718 were already available in the VLA archive. For NGC~3718, only the C-array data has been analyzed as the D-array data was lacking good phase calibrator measurement required for our purposes. \par
Flux calibrator measurements were performed at the beginning and at the end of each observation cycle. The phase calibrator was observed before and after each source cycle with a maximum distance between source and phase calibrator of 12$\degr$.
The correlator spectral setup used was spectral line mode 4 with Hanning smoothing and 64 channels per 1.5625 MHz channel width per frequency band providing a frequency resolution of 24.414 kHz/channel ($\sim$ 5.2 km s$^{-1}$). NGC~2782 was observed with a resolution of only 48.83 kHz ($\sim$ 10.48 km s$^{-1}$). Our observations have an average on-source integration time of 2.7h in the C-array and 2.8h in the D-array configuration. Thus, the corresponding RMS flux sensitivity of 0.6 and 0.8 mJy/beam/channel gives a 3-$\sigma$detection limit of $\sim 4 \times 10^{19}$ cm$^{-2}$ and $\sim 0.38 \times10^{19}$ cm$^{-2}$ column density for the C- and D-array, respectively. \par

The data reduction was performed following the standard procedure of the Astronomical Image Processing System \citep[AIPS;][]{gre03}. Calibration solutions were first derived for the continuum data-set (inner 3/4 of the spectral band width) and then transferred to the line data. The bandpass solutions were determined from the phase calibrator measurements to account for channel to channel variations in phase and amplitude.\par

After the calibration was applied, the two-IF data set was split into single-source data sets with 64 channels each. For each set the first 8 and last 8 channels were discarded, resulting in two 48-channel data sets. These were then combined into a final cube of 96 channels with an effective bandwith of 2343.7 kHz ($\sim$ 500 km s$^{-1}$) and a channel width of 24.4kHz ($\sim$ 5.2 km s$^{-1}$). For the THINGS data and NGC~2782 only the first 4 and last 4 channels were discarded which resulted in 112-channel sets with an effective bandwith of 2734.4 kHz and 5469kHz (NGC~2782), respectively. For NGC~1961 two independent frequency setups were used with an effective bandwith of 4687.5 kHz ($\sim$ 1018 km s$^{-1}$). After that, the calibrated C and D array data cubes were combined. \par

The task UVLIN within AIPS was used to subtract the continuum emission. The continuum emission was measured only in channels without line signal (usually the first 10 and last 10 channels of a data cube). The fit was done with a first order polynomial, namely a DC term (offset from zero) and a slope. For NGC~5953 only the first 8 channels were used (simple offset fit) because of additional line signal in the last channels.
Channels affected by galactic HI emission (NGC~4569, Channel 4-7) and internal RFI (NGC~1961, Channel 17-18) were dismissed and have been excluded from the further analysis. However, these channels were well outside line emission in both cases.\par

After the continuum substraction, the data were CLEANed using the task IMAGR within AIPS.
The calibration of each observation (C and D configuration) was verified individually (by a check of the imaged cubes). The flux-limit for CLEANing was set to a peak value in the residual image of 2$\times$RMS with a maximum iteration number of 10$^5$ steps. After CLEANing the residuals were rescaled by the ratio of the restoring beam area to the dirty beam area. 
The reason for this rescaling is that typically the residual flux of the source in cleaned channel maps is overestimated due to the different shape of the dirty and CLEAN beams. 
However, the results of the rescaling method turned out to not be significantly different from the results obtained without rescaling (less than 5\%).
We produced two CLEANed data cubes for each galaxy: 1.) Naturally weighted imaging 
with a velocity resolution of $\sim$20.8 km s$^{-1}$ and an average angular resolution of 33$\arcsec$, in order to obtain a datacube with maximum sensitivity 
to the outer HI disk and for HI companions around it.
2.) Robust weighted imaging for a high angular (20$\arcsec$) and spectral ($\sim$5.2 km s$^{-1}$) resolution data cube to study the galactic HI disk itself. 
To find the best compromise between angular resolution and RMS, several robust weighting parameters were tested and a robust weighting parameter of 0.3 was selected. The cellsize of each grid was set to 4$\arcsec$/pixel for natural and 2$\arcsec$/pixel for robust weighting with a resulting mapsize of 1024 $\times$ 1024 pixel. That corresponds to a field of view of $\sim68 \arcmin$ (natural) and $\sim34 \arcmin$ (robust) such that the 30$\arcmin$ primary beam is fully mapped in both cases. The RMS values and beam sizes for natural/robust weighting are listed in Table~\ref{tab_obs} with an average achieved RMS value of 0.34/0.46 Jy beam$^{-1}$ respectively. The channel maps for the naturally weighted cubes are presented in the appendix (see Fig.~\ref{fig_channel} as an example).
All further analysis was performed on these data cubes. A primary beam correction was only applied for the flux measurement.  \par

The subsequent analysis has been done with the Groningen Image Processing SYstem \citep[GIPSY;][]{Hulst}. The channel maps were combined to produce zeroth (intensity map), first (velocity field) and second (dispersion field) moments of the line profiles using the task MOMENT. The RMS values have been measured in two regions where no line signal was apparent and averaged over all channels of the cubes. A flux cut-off of three times the channel-averaged RMS value was used for the moment maps. In order to exclude one channel wide noise peaks from the moment maps, two adjacent channels with flux values above the cut-off level were required (corresponding to a width of 10.4 km s$^{-1}$ for the robust weighted cubes). 
However, for the natural weighted data cube we did not apply such a requirement since a single channel has already a width of 20.8 km s$^{-1}$, which is also the range of the minimal velocity width of an expected HI emission line.

\section{Analysis and Results}
\label{sec:res}
In the following the distribution and kinematic properties of the HI gas disks of the HI-NUGA sample will be derived, compared to optical properties of the host galaxies and analyzed as a function of the AGN type.

\subsection{HI Distribution and Environment}
\label{subsec:res_env}

\subsubsection{HI Gas Morphology}
The velocity-integrated HI intensity maps and the optical light distribution (DSS2-blue maps) are presented in Fig.~\ref{fig_mom0}. The HI intensity maps show a large variety of morphologies in the disk gas distribution including spiral (e.g. NGC~6951), ringed (e.g. NGC~7217) and centrally concentrated geometries (e.g. NGC~5953). A simple classification by eye has been used to assign a ring or spiral morphology. In cases where the distinction was unclear, the intensity map was projected into polar coordinates, giving the azimuthal variation of the intensity around a given center. If the intensity was constant along a radius, a ringed geometry was assumed. 
An overview of the HI-morphology of our sample is presented in Tab.~\ref{tab_comp}. In summary, we find spiral geometry in 8 galaxies, ringed geometry in 4 galaxies, and concentrated geometry in 5 galaxies, whereby NGC~4826 shows concentrated/spiral geometry and NGC~3368 ringed/spiral geometry in the central/outer disk. For NGC~3718 no classification is possible as its disk is nearly edge-on and shows in addition a large warped disk. \par

In the following, we describe the HI morphology of the disk for the three galaxies which clearly differ from the rest of the HI-NUGA sample:
NGC~5248 is the only galaxy of our sample which shows a large-scale HI "gas bar" (13 kpc in length). This HI bar coincides with a stellar bar (see left panel of Fig.~\ref{fig_mom0}). 
In addition, NGC~5248 shows two large HI spiral arms located outside the HI bar itself. Thus its HI morphology is remarkably similar to NGC~4151 which has a gas-rich bar and optically faint but gas-rich outer spiral arms \citep{Mun99}. \par

NGC~3368 exhibits two rings with radii of about 78 $\arcsec$ ($\sim$3.1 kpc) and 220 $\arcsec$ ($\sim$8.6 kpc) for the inner and outer ring, respectively. The outer ring could also be two closely wound spiral arms which are not resolved at our resolution of 19$\arcsec$. The inner ring is $\sim$15 times larger than the small-scale ( $r\approx5''$) nuclear "mini-bar" \citep[seen in the optical and NIR,][]{Moiseev}, which extends along a position angle of $120^\circ{-}125^\circ$. \par

NGC~4826 has two counter-rotating HI gas disks \citep{Braun} where the inner HI disk has a radius of $\sim$70 $\arcsec$ ($\sim$1.4 kpc) and the outer HI disk a radius of $\sim$700 $\arcsec$ ($\sim$13.9 kpc). No obvious asymmetries are seen, neither in the inner disk nor in the outer disk. Here the radius of the optical disk ($\sim$300$\arcsec$) lies between the radii of the two counter-rotating parts of the HI disk.\par 

Two galaxies in our sample show very distorted outer HI distributions in the form of HI tails and large faint extensions stretching from only one side of the main galaxy. NGC~2782 has a large HI tail with a projected length of $\sim 6 \arcmin$ \citep[$\sim$3 times the HI radius, see \ref{subsec:res_stellar} and][]{Smi91, Smi94} extending from its center to the north-west. NGC~1961 shows a large HI extension to the north-west with a projected length of $\sim 4 \arcmin$ \citep[$\sim$ 2 times the HI radius; see also][]{Sho82} from the center.
NGC~3718 exhibits a large HI warp \citep{Schwarz85}.\par 

Finally, two galaxies show large differences between the distribution of the stellar light and the HI gas: For NGC~6574 the HI gas distribution shows two HI peaks inside the stellar disk well separated by 30$\arcsec$ even at our angular resolution of 19$\arcsec$ (see Fig.~\ref{fig_mom0}). NGC5953 forms a close binary (likely merging) pair with NGC5943 (see left panel Fig 3) with the two galaxy centers separated by only ~60$\arcsec$ and a difference in systemic velocity of $\sim$53 km s$^{-1}$. The HI distribution shows a wing-like structure whose velocity field could be interpreted as a single disk, roughly uniformly distributed with a faint HI extension to the north. Also the HI velocity field shows no obvious tidal asymmetries which could be associated with the merging of the two optical galaxies.
Our study of the mass content (described in \S \ref{subsec:res_mass}) revealed a larger value for NGC~5953 (Log($M_{HI}/D^2)=7.1$) than the typical HI mass content \citep{Bet03} for a galaxy of that Hubble type (Log($M_{HI}/D^2)=6.5$). Hence an additional contribution of the gas content from NGC~5954 (Hubble type: Sc) might explain the gas surplus since the typical HI mass content for a Sc galaxies is Log($M_{HI}/D^2)=6.9$.  Thus we conclude that the ongoing merger already has changed the HI appearance of NGC~5953 and/or NGC~5954.

\subsubsection{HI Environment and Tidal Interactions}
In order to search for the presence of HI companions we used widefield maps (described in \S \ref{sec:obs}) with a field of view (FOV) of $\sim 60 \arcmin$ which are sensitive to companions with HI masses of $\gtrsim10^7 M_{\sun} \cdot d^2$ (lower detection limit near the center of the image) and $\gtrsim10^9 M_{\sun} \cdot d^2$ (at radial distances of 30$\arcmin$). Note that the sensitivity scales with the galaxy distance squared, indicated as $d$ in units of [10Mpc]. We found that 5 galaxies are surrounded by at least one HI-detected companion with a range in HI mass of (0.01 - 2.8)$\;10^9\;M_{\sun}$. In addition, potential optical companions were identified by using NED within our FOV (equivalent to a radial distance of 30$\arcmin$) and a velocity difference of $\pm$500 km s$^{-1}$ from the host galaxy ($\simeq$ 2 times the HI velocity bandwith). Optical companions (within these constrains) were found for $\sim$56\% of our galaxies. 
All cataloged optical companions (NED, within the constraints described above), their systemic velocity difference to the host galaxy and their distance from the host galaxy center are listed in Tab.~\ref{tab_sat}. We assume that the relative velocity between galaxy and companion is inversely proportional to the disruption time of the galaxy disk. For companions with associated HI gas detection (10 companions in total, $\equiv$ 38\% of the optical companions) also the derived HI-flux, HI-mass and their velocity width are provided. \par

To estimate the total tidal strength of companions on each galaxy, we used the formalism developed by Dahari (\citeyear{Dah84}). The strength of a tidal force affecting a galaxy with the mass $M_G$ and the diameter $D_G$  is primarily determined by the mass of the companion $M_c$ and the distance $R_{GC}$ to the power of three between the companion and the galaxy. We approximated the absolute distance by the projected distance and the mass of the companion by using the relation to its size, $M_C \varpropto D_C^{\gamma}$ with $\gamma=1.5$ \citep{Dah84, Ver07}. The tidal strength $Q_{GC}$ is defined as the ratio between the tidal force and binding force:
\begin{equation}
Q_{GC} \equiv \frac{F_{tidal}}{F_{bind}} \varpropto  \frac{\frac{M_c \times D_G}{R_{GC}^3}}{\frac{M_G}{D_G^2}} \varpropto \frac{(\sqrt{D_G D_C})^3}{R_{GC}^3}
\end{equation} 
The sum of the tidal interaction strengths (listed in Tab.~\ref{tab_sat}) created by all the companions in the field is then calculated as $\textbf{Q}=log(\sum_C Q_{GC})$. A value of $\textbf{Q}<0$ indicates that the tidal forces affecting the primary galaxy are smaller than the internal forces, and vice-versa. 
\par
Disturbances were characterized by using the divergence of the rotation curves between the approaching and receding side (see for example NGC~4321 in Fig.~\ref{fig_rot}) as well as obvious asymmetries in the HI intensity maps (see for example NGC~1961 in Fig.~\ref{fig_mom0}). A divergence of at least 5$\sigma$ (statistical error, roughly corresponding to 60 km s$^{-1}$) of the velocity difference (see \S \ref{subsec:res_kin}) in the outer disk between approaching and receding side was required (apart from deviations due to streaming motions along spiral arms). From these requirements we found that 7 galaxies have a disturbed outer HI disk (the individual galaxies are indicated in Tab.~\ref{tab_comp}). The origin of the disturbances (e.g. tidal forces, ram pressure) and the influence of the companions onto the HI disks are discussed in \S \ref{subsec:dis_env}

\subsection{Comparison of Stellar and Gaseous Distribution}
\label{subsec:res_stellar}

In order to compare the stellar distribution with the HI gas distribution, radial profiles were obtained by ellipse fitting using the task ELLINT within GIPSY. The radial profiles of the mean density and the cumulative sum of the optical light (from DSS2-blue images) and HI intensity maps were calculated. The widths of the radii were set to the HI angular resolution which is significantly lower than the typical angular resolution of $\sim 2\arcsec$ of the DSS2-blue images. In Fig.~\ref{fig_rad} the mean radial HI and optical profiles are presented (top panel) as well as the cumulative distribution of the HI and optical emission (bottom panel). \par
To allow for comparison of the spatial extent of the HI gas distribution with the stellar component, the ratio between the radius of the HI disk ($R_{HI}$) and the radius of the stellar disk ($R_{opt}$) is calculated. The optical radius $R_{opt}$ was set to the Holmberg radius defined as the parameter log D$_{B25}$ \footnote{Length of the projected major axis at the isophotal level 25 mag/arcsec$^2$ in the B-band} (Hyperleda). In order to obtain an equivalent radius $R_{HI}$ for the HI disk, we measured the radius at a fixed column density of $5.0\times 10^{19}$ cm$^{-2}$ for all our galaxies. The radius of the optical disk ($R_{opt}$) and the HI disk ($R_{HI}$) are indicated as a dashed and a solid line in Fig.~\ref{fig_rad}, respectively.
The results are listed in Tab.~\ref{tab_comp} for all galaxies and the mean ratio of $\frac{R_{HI}}R_{opt}{}$ is $\sim 1.7$. In addition the physical radii of the HI disks to this limiting column density are given in Tab.~\ref{tab_comp} with a mean absolute HI radius of $~20.6$ kpc.

\subsection{HI kinematics}
\label{subsec:res_kin}

\subsubsection{Kinematic Parameters and Rotation Curve}
The HI morphology provides detailed information about the spatial distribution of the gas, but only in combination with their kinematics can the complete dynamical phase space of the atomic gas be analyzed. Therefore we have performed a detailed kinematic analysis for all galaxies, which includes velocity fields, rotation curves, and residual velocity fields. 
Fig.\ref{fig_kin} shows (from the left to the right) the intensity-weighted HI mean velocity fields, the HI intensity (color) overlaid with the velocity field (contours), the HI velocity dispersion and the residual velocity field derived from a smooth fit to the rotation curve. \par

The following kinematic parameters (listed in Tab.~\ref{tab_kin}) have been derived from the observed velocity field (derived from the robust weighted cubes):
\begin{itemize}
\item The dynamical center and its offset from the optical center (taken from Hyperleda).
\item The systemic velocity (Heliocentric radial velocity) $v_{sys}$ in km s$^{-1}$.
\item The position angle (PA) in degrees, defined as the angle between the north-direction on the sky and the major axis of the receding half of the galaxy in anti-clockwise direction.
\item The inclination ($i$) in degrees.
\end{itemize}

These parameters were derived by fitting tilted rings to the velocity field using the task ROTCUR within GIPSY. Initial starting parameters have been taken from cataloged data (NED or Hyperleda). For the fit we excluded data points within an angle of 20$\degr$ of the minor axis. The widths of the radii were set equivalent to the corresponding angular resolution of each galaxy (see Tab.~\ref{tab_obs}). No radial velocity component was fitted. The kinematic parameters have been derived in an iterative way following Begeman (\citeyear{Beg89}). The parameters were assumed to be the same at all radii (unless specified otherwise), except for the circular velocity. 
Therefore we weighted the obtained value in each ring with its standard deviations, in order to obtain the mean parameter.
The iterative process of deriving the kinematic parameters is as follows:
we began by fitting both sides of the velocity field in order to determine the systemic velocity (with $PA$, $i$ held fixed, everything else was free to vary), then the kinematic center (with a fixed $PA$, $i$, $v_{sys}$) and checked again the systemic velocity (with a fixed $PA$, $i$ and center).  Next, the derived center, the inclination and the systemic velocity were held fixed in order to determine the position angle. The position angle was well constrained by the model fits to within $\pm$(1-5)$\degr$. The center, the position angle and systemic velocity were held fixed in order to obtain the inclination angle. In case of insufficient precision (statistical error of the inclination fit was $\geq 8\degr$) for the derived inclination (for NGC~2782, NGC~4569, NGC~6574) we applied an ellipse fit to the optical image. The inclination derived from the optical axial ratio measured at the Holmberg radius was then used to fit the rotation velocity. 
Finally, the rotation velocities were obtained with fixed inclination, position angle, center and systemic velocity. We used for NGC~4826 two different PA (inner part: R=0 to 64$\arcsec$; outer part: R=64 to 700$\arcsec$)  to account for its counter-rotating HI disk. For NGC~3718, the inclination was required to increase from $i$=48$\degr$ to 60$\degr$ to model its warped HI disk. The derived PA was constant over the disk within $\pm5\degr$. The derived rotation curves for each galaxy are presented in Fig.~\ref{fig_rot} for a) the entire disk (black line), b) the approaching side of the disk only (dark grey dots), and c) the receding side of the disk only (light grey dots). \par

The derived kinematic parameters and rotation curves have been verified by using INSPECTOR in GIPSY. This task allows us to overlay the circular HI velocities on position-velocity (pv) diagrams in order to verify the tilted ring parameters on the 3-dimensional datacube. Comparison between the derived rotation curves and corresponding pv slices from the data cubes (center and right panel of Fig.~\ref{fig_rot}) show the effect of beam smearing (particularly in the inner 20$\arcsec$) and some small misalignments ($\sim$ spatial and spectral resolution) between our modeled parameters and observed kinematics (e.g. center, systemic velocity). Otherwise, the correspondence between model and data is good in all cases (see next subsection).

\subsubsection{Velocity Fields and Velocity Dispersion}
2-D model velocity fields have been created from the derived interpolated rotation curves using the program \textit{VELFI} within GIPSY. 
These model velocity fields were subtracted from the observed velocity fields to obtain residual velocity fields. The typical range of the residual velocities is $\pm$ 25 km s$^{-1}$. Furthermore, the residual velocity fields (Fig.~\ref{fig_kin}) can also be utilized to identify systematic misalignments of the derived kinematic parameters \citep{Warner}. However, no obvious misalignments were identified.\par

Most of the galaxies in our sample exhibit fairly regular velocity fields that are dominated by circular motion. The three exceptional cases are described in the following: NGC~2782 exhibits a large gaseous tidal tail (see \S \ref{subsec:res_env}) with a velocity gradient of $\sim$100 km s$^{-1}$ over $\sim$65kpc (360$\arcsec$) across that tail. The residual velocity field shows that the HI kinematics in the inner HI disk (which coincides with the stellar disk) are fairly different. Smith (\citeyear{Smi91,Smi94}) using HI data at $\sim$9$\arcsec$ resolution found in addition 1) a shorter HI structure extending toward the east (with 6.6$\cdot 10^8 M_{\sun}$) and 2) that the HI gas in the inner disk is counterrotation with respect to the tidal material, suggesting a merging of two galaxies of unequal masses. Due to the resolution of 34$\arcsec$ of our HI data, this disk is basically unresolved and the derived rotation curve reflects partially the kinematic properties of the tidal tails.

NGC~5953 exhibits significant streaming motions, inconsistent with global rotation, with a velocity gradient of $\sim$60 km s$^{-1}$ over 100$\arcsec$ ($\sim$17kpc). Also NGC~1961 shows streaming motions in the northern part of its HI distribution (velocity gradient $\sim$400 km s$^{-1}$ over 180$\arcsec$, $\sim$47kpc), in addition to a rotating disk which is coinciding with the optical one.
The residual velocity fields revealed for some galaxies of our sample (e.g. NGC~5248, inner disk of NGC~6951) indications for non-circular motions. Such motions can be caused by a bar potential or spiral waves \citep{Kruit}.\par
The velocity dispersion maps show that the majority of our galaxies have an average velocity dispersion between 10 and 30 km s$^{-1}$. The reason for the larger dispersion along the minor axis seen in almost all galaxies is likely caused by an observational effect: the angular distance between adjacent velocities becomes smaller along the minor axis, and hence the beam covers a larger velocity gradient. For NGC~3147, NGC~6574, NGC~3627 and NGC~4569 we found slightly higher dispersion with an average of 15-25 km s$^{-1}$ and maxima of 40-70 km s$^{-1}$. Further, the highly-disturbed galaxies (NGC~1961, NGC~2782, NGC~5953) exhibit very large velocity dispersion with maxima of 60-100 km s$^{-1}$.

\subsection{Gaseous and Dynamical Masses}        
\label{subsec:res_mass}

The HI fluxes of the galactic disks were measured in the intensity maps of the naturally weighted cubes. The intensity maps were corrected for the response of the VLA primary beam. In Tab.~\ref{tab_mass} the HI flux and the corresponding HI mass are listed for each galaxy. 
The average HI mass of our sample is $5.4 \times 10^9M_{\sun}$ with a maximum mass of 46.6$ \times 10^9M_{\sun}$ (NGC~1961). \par

To obtain the dynamical mass we used two approaches. A Brandt curve \citep{Brandt} was fitted (black curve) to the measured rotation velocities of the whole disk. The Brandt curve is an approximation of a solid-body rotation at small radii and a Keplerian-type velocity decrease at large radii and can be described by (see equation [26] and [28] of Brandt)
\begin{equation}
\frac{v(R)}{v^{Br}}=\frac{R}{R^{Br}} \left[ 1/3+2/3 \biggl(\frac{R}{R^{Br}}\biggr)^n \right]^{-3/2n} ,
\end{equation} 
where $v^{Br}$ is the maximum rotation velocity at the radius $R^{Br}$ and $n$ is an index parameterizing the steepness of the curve. From the derived parameters the dynamical mass $M_{dyn}^{Br}$ can be calculated (equation [29] of Brandt) via
\begin{equation}
M_{dyn}^{Br}=1127 \; \biggl(\frac{3}{2}\biggr)^{3/n} R^{Br} \; v^{Br} \; D
\end{equation} 
with $M_{dyn}^{Br}$ in solar mass units, the distance $D$ in [Mpc], the velocity parameter $v^{Br}$ in [km s$^{-1}$] and the radius parameter  $R^{Br}$ in [\arcsec]. For all galaxies the derived dynamical masses are listed in Table~\ref{tab_mass}. In addition, we have calculated the dynamical mass via the spherical symmetry approximation:
\begin{equation}
M_{dyn}=\frac{R \; v^2}{G} , \quad M_{dyn}[M_{\sun}]=232407 \; R_{HI}[kpc] \; v^2[(km \; s^{-1})^2],
\label{eq_mass}
\end{equation} 
with the circular velocity $v$ from the rotation curve at the HI radius $R_{HI}$ (see \S \ref{subsec:res_stellar}). Note that the mass could be 40\% lower, if a disk geometry would be assumed instead of a spherical one. 
The dynamical mass $M_{dyn}^{Br}$ obtained by the Brandt fit and the dynamical mass derived at the HI radius $M_{dyn}$ are presented in Tab.~\ref{tab_mass}. The comparison between both methods shows that the mass derived by the Brandt fit is on average a factor of two larger as the mass derived at the HI radius. This difference might be expected since the Brandt method is based on the assumption that the rotation curve reaches a Keplerian fall off quite rapidly, which is not exactly the case for most of our galaxies (probably due to dark matter contribution in the outer disk regions). Hence we will use in the following calculations of this paper the dynamical mass derived from Eq.~\ref{eq_mass}. The ratio of the HI mass to the dynamical mass of a galaxy is given by $M_{HI}/M_{dyn}$ which lies for our sample in a range of 0.003 to 0.046 with an average ratio of 0.015. \par

\subsection{HI properties as a function of AGN type}
\label{subsec:res_agn}
Here, we investigate possible correlations between the observed HI properties of our galaxies and their type of nuclear activity. Since our sample, as described in \S \ref{subsec:obs_sample}, consists mainly of Seyfert (Sy) and LINER (L) host galaxies, we will focus on a comparison of these two classes. A summary of the average properties for each AGN type is listed in Tab.~\ref{tab_agn}. The error calculation is based on the bootstrapping method \citep{Efron} by taking the standard deviation of the mean values from 1000 replications.

\subsubsection{Environment and HI morphology}
As already described in \S \ref{subsec:res_env} we found in our sample several galaxies which are surrounded by companions and/or have disturbed outer disks. No significant statistical relation is found between HI environment and AGN type: The percentage of optical companions is roughly the same for each AGN-type (Sy: 3/7, L: 4/7), and the number of galaxies with HI-detected companions (Sy: 1/7 ; L: 2/7; HII: 1/2) is also similar given our small number statistics.
Galaxies with disturbed outer disks seem to be more abundant in LINER hosts (5/7) than in Seyfert hosts (1/7). The origin of these disturbances will be discussed in \S \ref{subsec:dis_env}. \par
In order to search for a possible correlation between the HI morphology and the AGN type, we excluded the following galaxies from the analysis, as their HI disk morphology is highly asymmetric and hence their morphology can not reliably be assigned:  NGC~1961 (large gas extention to NW), NGC~5953 (interacting system NGC~5953/54), NGC~3718 (warped HI disk with inner disk close to edge-on), NGC~4826 (very concentrated in center, counter-rotating HI disks). Thus the number of Seyferts and LINERs is reduced to 5 galaxies for each type. 
Only LINER host galaxies show ringed HI morphology (LINERs: 80\%, Sy: 0\%), while gas spiral arms appear as frequently in both types (Sy: 60\%, L: 60\%).
We used the bootstrapping method \citep{Efron} to calculate the statistical significance of this result: The confidence interval (at percentiles of [5, 95]\%) for LINERs is [40, 100]\% for ringed structure. The standard deviation of the mean values for HI rings is found to be (80$\pm$17)\%. \\
Furthermore, the radii of the HI rings were determined but neither the absolute radius was found to be the same for all ringed galaxies (range between 0.9-5.7 kpc), nor the relative radius of the ring  to the optical disk ($R_{ring}/R_{opt}$: 0.15-0.6) or to the HI disk ($R_{ring}/R_{opt}$: 0.11-0.6) shows any obvious trend.    

\subsubsection{Masses}
In order to search for a possible relation between the mass properties and the nuclear activity type, we compared the dynamical mass (within the HI radius $R_{HI}$) as well as the HI mass as a function of the AGN type. The mean dynamical mass is slightly larger for LINERs (3.5$\cdot10^{11} M_{\sun}$) than for Seyfert galaxies ($M_{dyn}$ is 2.4$\cdot10^{11} M_{\sun}$) but similar within a range of [0.45-5.7]$\cdot10^{11} M_{\sun}$ (Sy) and [0.4-12.3]$\cdot10^{11} M_{\sun}$ (L). For the HI gas content, we found for Seyfert galaxies a smaller mean HI mass (2.5$\cdot10^{9} M_{\sun}$) than for LINERs (8.6$\cdot10^{9} M_{\sun}$) with a range of [0.2-9.0]$\cdot10^{9}M_{\sun}$ and [3.8-46.6] $\cdot10^{9}M_{\sun}$, respectively. In order to study the relative gas content we calculated the ratio of $M_{HI}/M_{dyn}$ for each galaxy and averaged it for a given AGN type. The mean ratio of $M_{HI}/M_{dyn}$ is 0.009 for Seyfert and 0.013 for LINER host galaxies. Interestingly, the two non-AGN galaxies in our sample have a larger HI gas content of 0.046 and 0.028. The mean Hubble type is roughly the same for our Seyfert ($t$=2.8), LINER ($t$=2.7), and non-AGN ($t$=2.6) sample.\par 

A comparison between the relative gas content ($M_{HI}/D_{25}$) versus Hubble type for our samples (Sy, L, non-AGN) and typical values for a larger sample \citep{Bet03} is presented in Fig~\ref{fig_Hubble}. On average the relative gas content of the sample of Bettoni et al. seems to be larger compared to our sample. Note that Bettoni et al. (\citeyear{Bet03}) excluded galaxies cataloged as having distorted morphology and/or any signature of peculiar kinematics (such as polar rings, counterrotating disks or other kinematically decoupled components) and thus their sample does not contain galaxies similar to ours.

\subsubsection{HI versus optical extent}

In order to compare the relative gas disk size between different AGN types, we averaged the ratio $R_{HI}/R_{opt}$ of the HI radius $R_{HI}$ to the optical radius $R_{opt}$ (as described in \S \ref{subsec:res_stellar}) for each AGN type. This analysis shows that the gas disk in Seyfert galaxies is slightly larger with an average ratio of 1.9 for Seyferts than for LINERs (1.4). However, the difference is within the statistical deviation ($\pm \sim$0.6). If we neglect galaxies which are part of a galaxy group or cluster (NGC3627, NGC4569, NGC4579), the ratios increase to 2.4 and 1.5, respectively (which corresponds to a difference of 0.9 $\pm$ 0.6). Thus, there is a hint that Seyfert galaxies might eventually have more extended HI gas disks than LINER galaxies. \par

\subsection{Synopsis of results}

The atomic gas distribution and kinematics for a sample of 16 galaxies have been presented. There is a large variety of HI distribution in the sample (spiral-, ringed- and centrally concentrated geometries) and some galaxies show strongly disturbed HI disks. Most of our galaxies exhibit fairly regular HI gas distribution, exceptions are discussed below. The analysis of the large environment revealed that 5 galaxies have HI-rich companions and for several galaxies potential optical companions are present. The distributions of the gaseous and stellar component are compared, kinematic properties of the HI gas are analyzed (i.e. rotation curves, residual velocity maps), and the gaseous and dynamical masses are derived. Furthermore, a comparison of HI properties and AGN type suggest correlations with HI morphology and host disturbance: LINER host galaxies seem to have a larger fraction of disturbed HI disks, a significant higher percentage of HI ring geometry, and tend to possess a smaller HI extent (relative to the optical radius) than Seyfert host galaxies. No correlation between the presence of companions and the AGN type was found.

\section{Discussion}
\label{sec:dis}

The relationship between large scale environment and nuclear activity has been long debated in the literature. Several studies have found indications for correlations between the environment and nuclear activity \citep{Sto01, Cha02, Mar03}. In particular it was suggested that interacting galaxies or galaxies with companions exhibit a significant excess of nuclear activity compared to isolated galaxies \citep{Dah84, Kee85, Raf95}. On the other hand no relation was seen by other studies \citep{Vir00, Sch01, Fue88, Mac89, Lau95}. Hence the issue of possible relations between the environment and nuclear activity appears to be still controversial. 
Furthermore, results from Keel et al. (\citeyear{Kee85}) suggested that nuclear phenomena might likely be triggered by a tidally induced inflow of gas from the disk to the nuclear regions, rather than gas transfer between the interacting galaxies themselves. Keel (\citeyear{Kee96}) also  found that Seyfert galaxies in pairs actually display smaller kinematic disturbances than non-Seyfert galaxies in pairs, which is obviously in disagreement with the hypothesis that tidal interactions are necessary for the transport of angular momentum and the fueling of the SMBH. 
Since most of these studies are based on optical/IR imaging, they are in principle less sensitive to distortions than our study of the atomic gas that reacts most readily to tidal disturbances. It should be pointed out that a detailed study of HI gas properties for active versus non-active galaxies (Mundell et al., in prep.) is under-way. \par 

In the context of different AGN types, the analysis of a sample of 451 active galaxies (Sy, LINER, Transition, HII, and absorption-line galaxies) from the Palomar survey \citep{Sch01} showed no correlation between AGN-type and the percentage of galaxies with nearby companions after taking morphological differences of the host galaxies into account. We also see no evidence for a correlation between the fraction of companions and the AGN type present (Seyfert, LINER galaxies), neither from our HI study nor for cataloged optical companions listed in NED (see \S \ref{subsec:res_env}). Note that our sample has a limited number of LINERs (7 galaxies) and Seyfert galaxies (7) and hence the existence of possible relations can not be completely excluded. \par

\subsection{HI environment:  tidal forces and their correlation with disturbed HI disks}
\label{subsec:dis_env}
Companion galaxies can possibly disturb the HI gas in a galactic disk via tidal forces and hence affect the fueling of the center with gas \citep[e.g.][]{Kee85}. The sensitivity and endurance of HI to trace the strength and prevalence of tidal interactions among Seyfert galaxies is discussed in detail by Greene et al. (\citeyear{Gre03}). 
In our analysis we found companions for about half of our sample. In almost all cases the produced tidal forces are smaller than the binding forces of the affected host galaxy (indicated as $\textbf{Q}<0$, see \S \ref{subsec:res_env} and Tab.~\ref{tab_sat}).

Only the system NGC5953/54 exhibits signs for very strong gravitational interaction as mentioned in \S \ref{subsec:res_env}. 
Therefore, most of the disturbances in our sample (6/7 galaxies) can not be explained simply by tidal forces presently at work. 
The most probable explanations for these disturbances are described in the following, listed in decreasing relevance:
\begin{itemize}
\item Interaction with a companion now further or far away. If we assume a relaxation timescale of the disturbance of 3$\cdot 10^8$yr (typical time for one rotation of a galaxy) and a maximum fly-by velocity of $\sim$ 500 km s$^{-1}$, the involved companion is expected to be found within a radial distance of $\sim$ 150 kpc from the disturbed galaxy. Since 71\% of the disturbed disks show companions in a reasonable distance (projected distance is less than 150 kpc), tidal interactions in the past seem to be primarily responsible of the disturbances identified in the HI disks.
\item Ram pressure stripping \citep{Cay90, Vol00, Vol01}. In particular for NGC~1961 stripping by intergalactic material was suggested \citep{Sho82}. But since none of our galaxies with disturbed HI disk lie in a massive group or cluster environment, gas stripping due to ram pressure is not very likely to explain the presence of disturbances in our sample where no nearby companion is found. Furthermore, the outer HI disk, where disturbances are usually seen, has been removed in clusters due to ram pressure stripping, indicated also by a small HI radius (e.g. NGC~4569, NGC~4579 as part of the Virgo cluster; see \S\ref{subsec:dis_stellar})  
\item Minor merging, whereby the companion has now fully merged and has left no optical trace. This might be the case for NGC~4736 and NGC~2782 where no companions are found in a reasonable distance for tidal interaction.
\item Large gas accretion from cosmic filaments: Asymmetries in the gas accretion may cause disturbances \citep[for effects of gas accretion on spiral disk dynamics see][]{Bou02}. 
\end{itemize}

\subsection{HI morphology and comparison with the stellar distribution}
\label{subsec:dis_stellar}

The comparison of the radial density profiles between the HI gas distribution and the stellar distribution revealed significant deviations as expected: 1.) The extent of the HI disk is on average 1.7 times the optical radius, indicated by the Holmberg radius. Only NGC~3627, NGC~4569 and NGC~4579 show a smaller HI radius than optical radius. This can be explained by the fact that they are all members of interacting groups and/or by their rapid motion through an intracluster medium. NGC~3627 is part of the Leo Triplet group and the past encounter with NGC~3628 could explain the spatial conincidence of both the stars and the gas \citep{Zha93}. The truncated disks of NGC~4569 (and NGC~4579) are most probably a signature of strong ram pressure stripping in the past by the intracluster medium which pervades the Virgo Cluster \citep{Cay94}. However, most of the galaxies in our sample show a larger HI disk than their optical one. 
2.) The radial density profiles exhibit for most of our galaxies a deficiency of HI gas in the inner part of the galaxy disk. This is in general explained by the phase transition from atomic to molecular gas in neutral ISM \citep{Young}.

\subsection{Correlations between HI gas properties and AGN activity type}
\label{subsec:dis_correl}
As described in \S \ref{subsec:res_agn} our analysis revealed that the number of galaxies with disturbed HI disks is higher for LINER galaxies than for Seyfert galaxies. But since several mechanisms which can not easily be distinguished can cause these asymmetries (see \S \ref{subsec:dis_env}), it is not possible to draw any strong conclusions explaining the higher fraction for LINERs. \par

Our study of the HI morphology revealed a significantly higher percentage of galaxies with a HI gas ring for LINER than for Seyfert galaxies. Interestingly, only the study of the Extended 12$\mu$m Galaxy Sample \citep{Hunt99b} indicated a prevalence of stellar rings in active galaxies (Seyfert and LINERs), where LINERs have elevated rates of inner rings, while the Seyfert host galaxies have outer ring fractions several times those in normal galaxies. However, we found no HI gas rings in Seyfert host galaxies. Note that stellar rings are not preferentially found in Seyfert, LINERs, or non-AGNs for our sample by using the optical classification from RC3 listed in NED (see classification in Tab.~\ref{table_intro}; indicated as R or r). To summarize, HI rings are more often found in LINERs while no strong correlation with stellar rings is present. \par

One possible explanation for an abundance of HI gas rings in LINERs may be a common evolution of the gas distribution in the disk together with the nuclear activity where both are subject to the influence of a present bar or previous one which has now dissolved. This becomes important since rings are linked observationally and phenomenologically to barred galaxy dynamics \citep[for a general review see][]{But96}, and  hence, expected to be seen after the bar had enough time to redistribute the gas toward the end of its life-time. Thus, a time evolution of AGN types seems to be possible, where Seyfert and LINERs represent different phases of the galaxy activity cycle: Seyfert galaxies are the ones where the fueling process has just been triggered (e.g. through disturbances and/or bar dynamics) while LINERs are the ones where the triggering mechanism has already distributed the gas in a more stable new configuration (rings) that does no longer support the massive inflow of gas. Interestingly, disturbances, which are assumed to be a sign for a recent trigger of gas inflow, are preferentially found in LINERs in our sample. That would suggest that LINERS are the first stage of activity, just after the triggering: The gas is still distributed in rings, and the fueling of the AGN has just started by the redistribution of the gas. However, as most of the disturbances in our sample are probably due to tidal interactions in the past, as explained in \S \ref{subsec:dis_env}, the currently observed nuclear activity must not be identical to the one during the ongoing tidal interaction. \par

A general scenario for self-regulated activity in low-luminosity AGNs was developed by \cite{Gar05}, in which the onset of nuclear activity is explained as a recurrent phase during the typical lifetime of any galaxy. In this scenario the activity in galaxies is related to that of bar instabilities, expecting that the active phases are not necessarily coincident with the phase where the bar has its maximum strength. Since the infall of gas driven by a bar is self-destructive \citep{Bou02}, i.e. it weakens and destroys the bar, the potential returns to axisymmetry and the gas piles up in a stable configuration (i.e. in rings at the resonances). At this stage, torques exerted by the gravitational potential are negligible and other competing mechanisms, e.g. viscous torques, must transport the gas in the center of the galaxy.
The periods of Seyfert/LINER activity (each lasting $\sim 10^8$ years) are suggested to appear during different evolutionary stages of a bar episode (typically characterized by a lifetime $\sim 10^9$ years), depending on the competition between viscosity and gravity torques. The prevalence of HI rings in LINERS, derived  in this work, could be explained by the scenario proposed by \cite{Gar05}, where AGN activity is linked to the evolutionary state of the bar-induced gas flow.  However, to further substantiate this link requires deriving the complete gravity torque budget based on the HI distribution (see Haan et al. in prep.).

Regarding the relative HI radius ($R_{HI}$/$R_{optical}$) we found a slightly larger HI extent for Seyfert galaxies than for LINERs, which increases even more when neglecting galaxies which lie in galaxy clusters or groups. 
In contrast, our study of the large environment, the HI gas content (by using the ratio $M_{HI}$/$M_{dyn}$), and the relative HI mass ($M_{HI}$/$M_{dyn}$) revealed no significant correlation with the AGN-type. This might be due to the limited number of galaxies in our sample. Hence we conclude that a detailed HI study with a larger sample may reveal more additional information on the interplay between the gaseous component and the AGN type present.

\section{Summary}
\label{sec:sum}
We present results on the atomic gas distribution and kinematics of 16 nearby spiral galaxies. This sample forms the part of the NUGA survey which studies the neutral (atomic and molecular) gas dynamics of nearby low luminosity AGN galaxies. 
We investigated relations between the atomic gas properties, the environment and the stellar distribution as well as possible relations to the AGN type present (Seyfert, LINER). Several galaxies are surrounded by companions with associated HI emission and half of our sample has dynamically disturbed disks.
No evidence for a correlation between the presence of companions and Seyfert and LINER galaxies is found, neither from our HI study nor for cataloged optical companions.\par

The HI morphology of our galaxies revealed rings, spiral arms, and centrally concentrated peaks. The comparison with the AGN type present showed that ring structure is significantly more often present in LINER than in Seyfert host galaxies. Since bars are dynamically linked to the presence of gas rings, a time evolution of the AGN activity together with the redistribution of the neutral gas seems to be a plausible explanation for this finding. Additionally, we found a slightly larger HI extent and less prevalent disturbed HI disks for Seyfert hosts than for LINER host galaxies. A detailed study with a larger HI sample may reveal more relations with respect to the AGN type and would improve the significance of our results. \par

\acknowledgments
The authors wish to thank the anonymous referee for helpful comments. We are grateful to the National Radio Astronomy Observatory (NRAO) for their support during this project. The NRAO is operated by Associated Universities, Inc., under cooperative agreement with the National Science Foundation. We acknowledge the usage of the HyperLeda database (http://leda.univ-lyon1.fr); \citep{Paturel} and the NASA/IPAC Extragalactic Database (NED) which is operated by the Jet Propulsion Laboratory, California Institute of Technology, under contract with the National Aeronautics and Space Administration. S.H. is supported by the German DFG under grant number SCHI 536/2-1.


\newpage


\begin{figure}[ht]
\begin{center}
\includegraphics[scale=0.8]{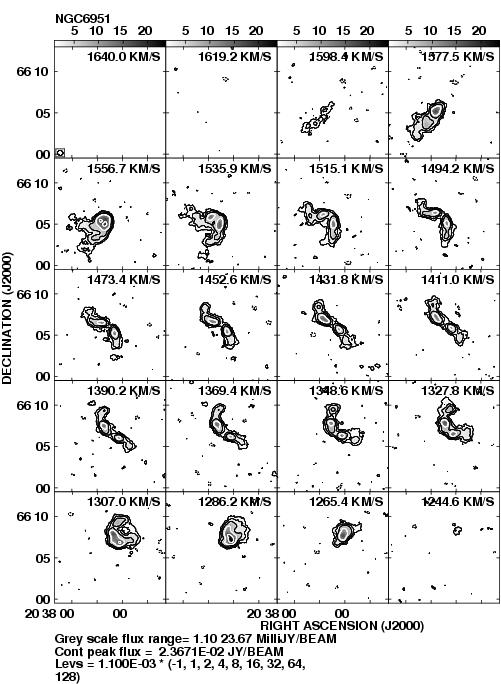}
\caption{
Channel maps of the natural weighted data cube for the example NGC~6951. Each channel has a velocity width of $\sim$20.8 km s$^{-1}$. The flux cut-off was set at 3$\sigma$ of the RMS value (for NGC~6951: 0.0011 Jy/Beam) and the contours are in steps of $3\sigma \cdot 2^n$Jy (n=1,2,3...). The channelmaps for all galaxies are available as online data (see Appendix).}
\label{fig_channel}
\end{center}
\end{figure}

\clearpage

\begin{figure}[ht]
\begin{minipage}[c][21cm][t]{5.4cm}
\includegraphics[scale=0.31]{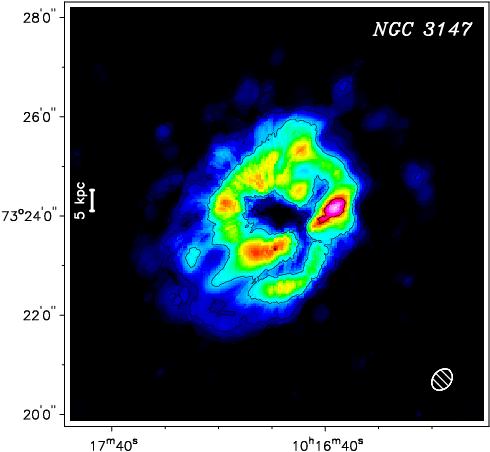}\\
\includegraphics[scale=0.31]{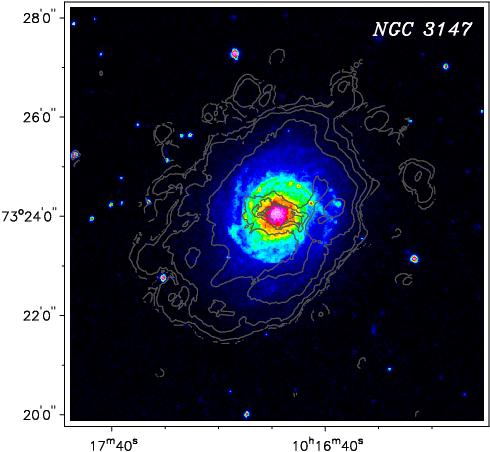}\\
\includegraphics[scale=0.31]{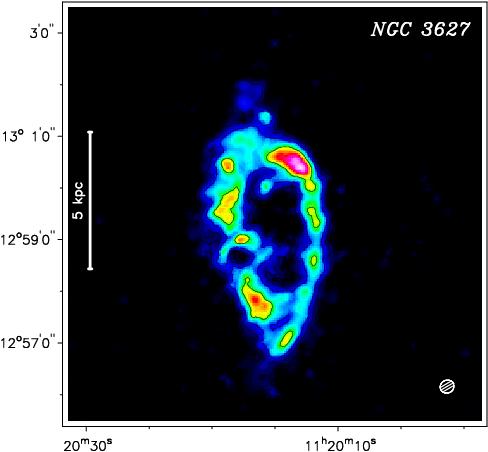}\\
\includegraphics[scale=0.31]{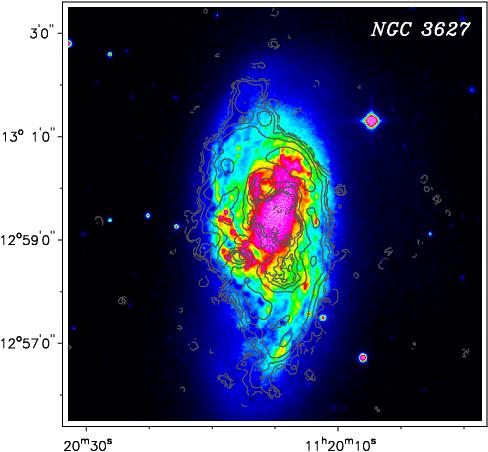}\\
\end{minipage} 
\begin{minipage}[c][21cm][t]{5.4cm}
\includegraphics[scale=0.31]{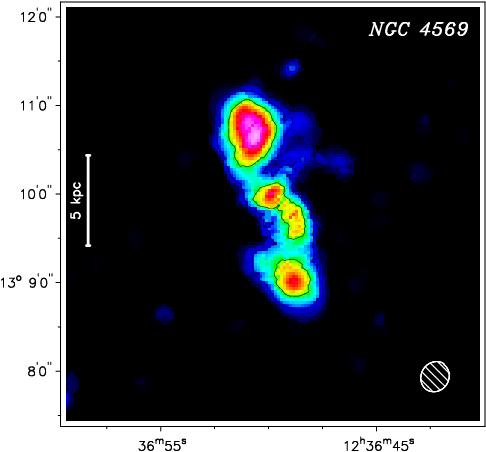}\\
\includegraphics[scale=0.31]{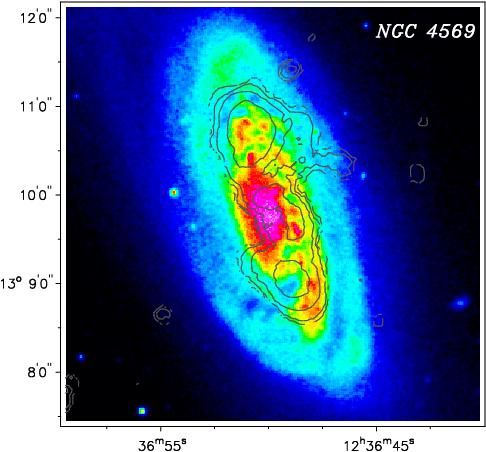}\\
\includegraphics[scale=0.31]{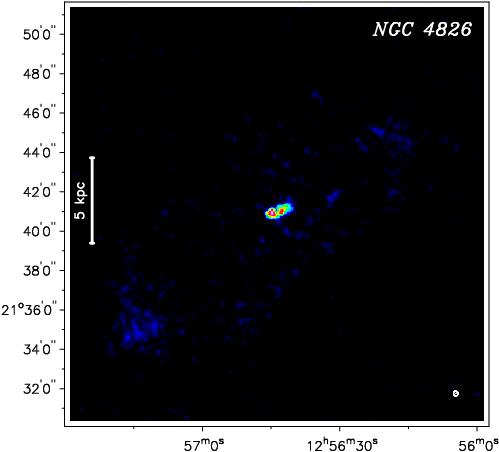}\\
\includegraphics[scale=0.31]{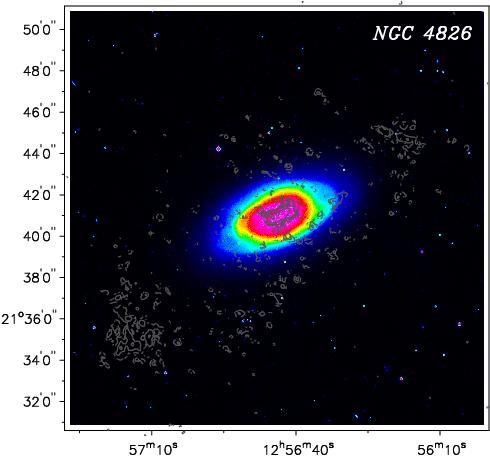}\\
\end{minipage} 
\begin{minipage}[c][21cm][t]{5.4cm}
\includegraphics[scale=0.31]{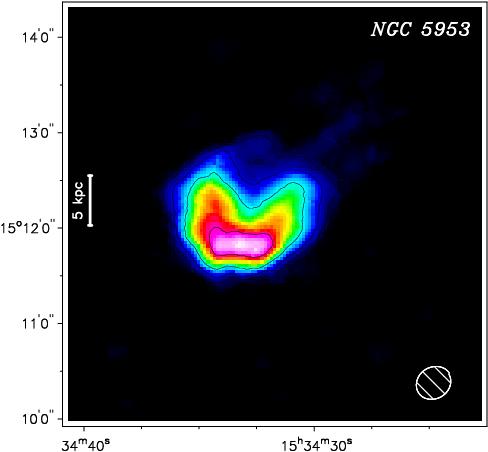}\\
\includegraphics[scale=0.31]{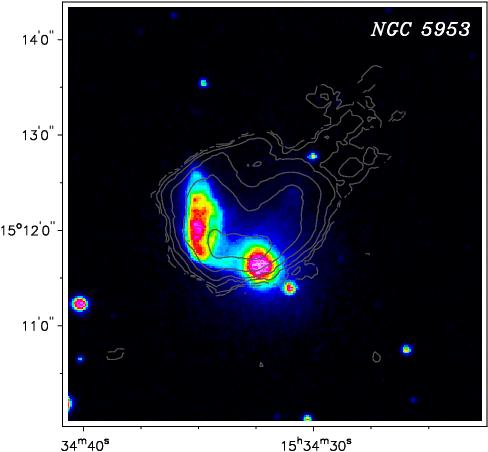}\\
\includegraphics[scale=0.31]{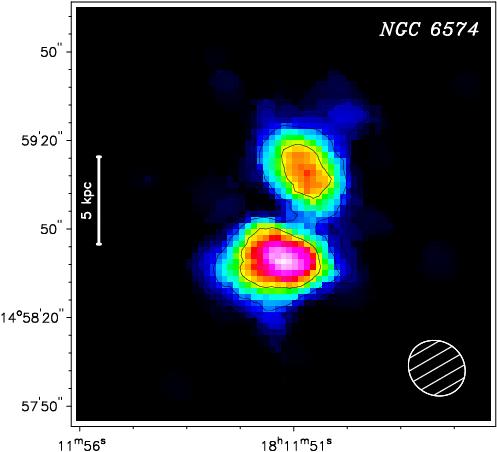}\\
\includegraphics[scale=0.31]{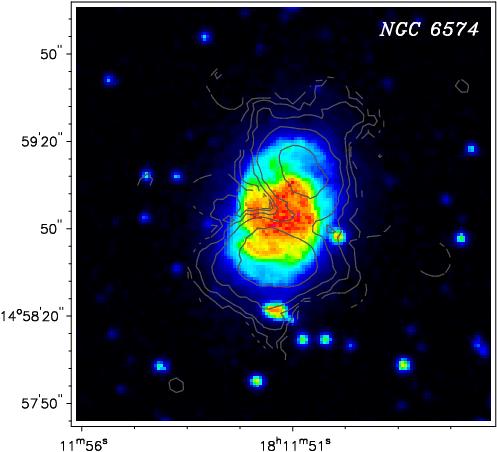}\\
\end{minipage} 
\end{figure}
\newpage
\begin{figure}[ht]
\begin{minipage}[c][21cm][t]{5.4cm}
\includegraphics[scale=0.31]{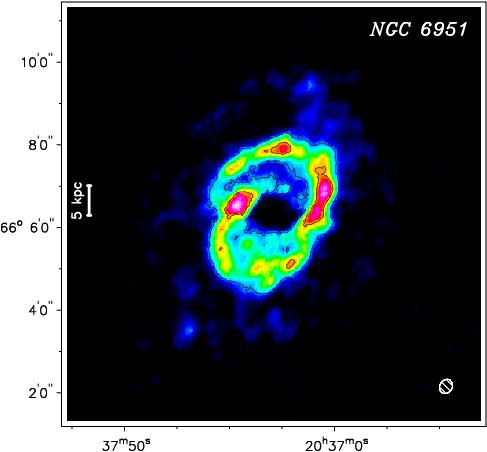}\\
\includegraphics[scale=0.31]{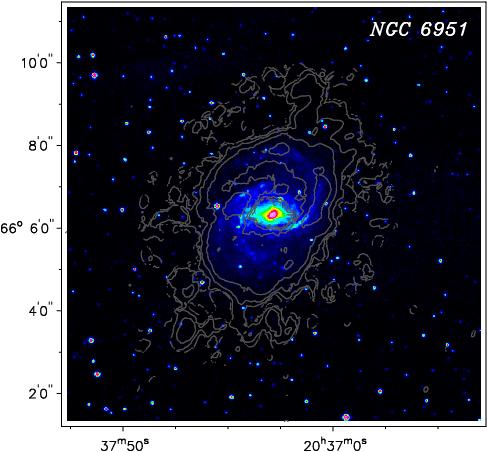}\\

\end{minipage} 
\begin{minipage}[c][21cm][t]{5.4cm}
\includegraphics[scale=0.31]{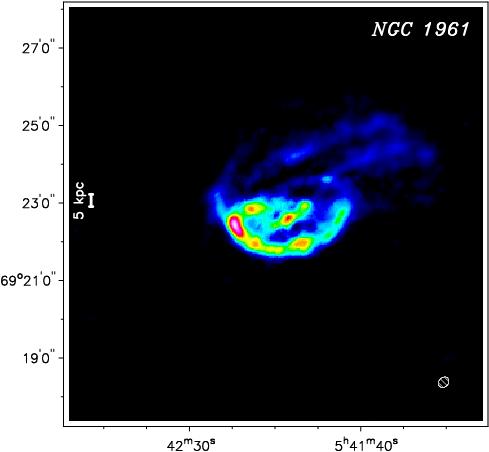}\\
\includegraphics[scale=0.31]{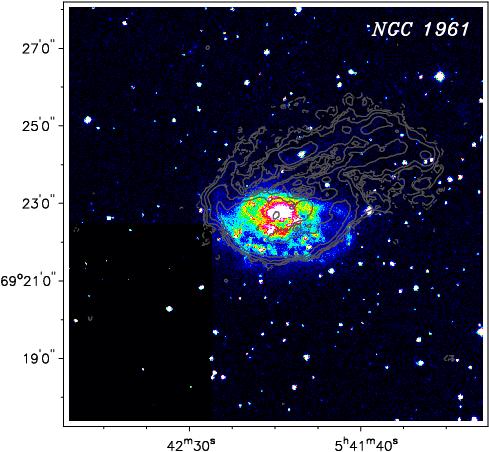}\\
\includegraphics[scale=0.31]{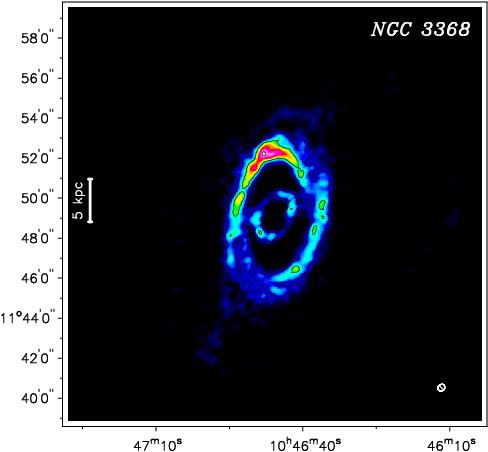}\\
\includegraphics[scale=0.31]{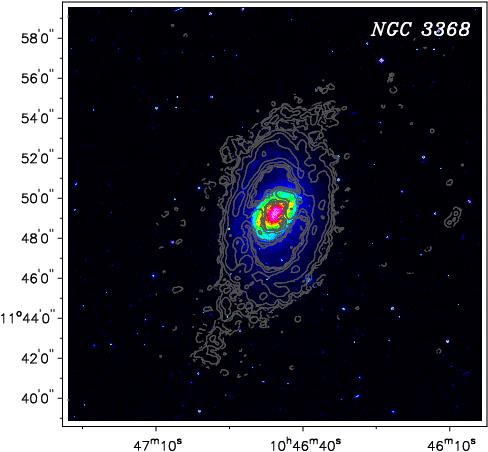}\\
\end{minipage} 
\begin{minipage}[c][21cm][t]{5.4cm}
\includegraphics[scale=0.31]{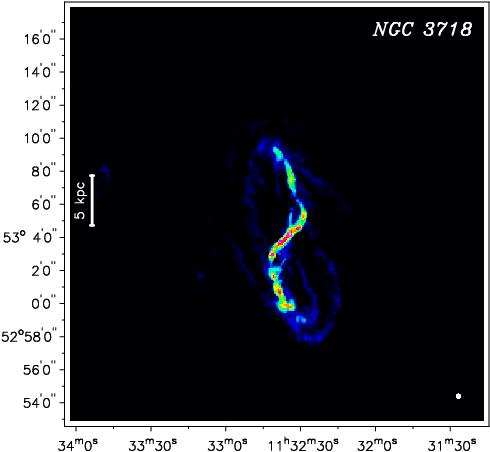}\\
\includegraphics[scale=0.31]{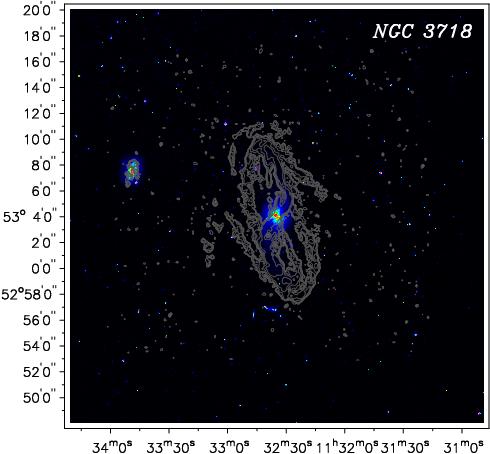}\\
\includegraphics[scale=0.31]{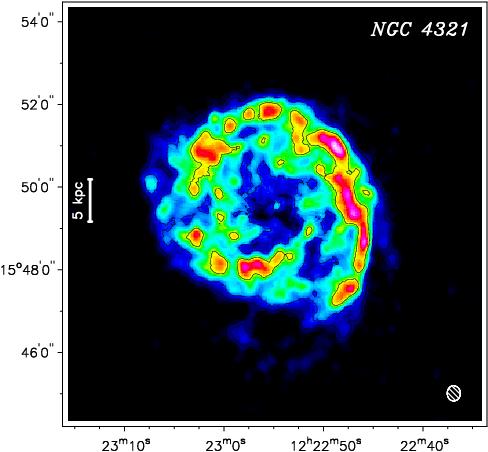}\\
\includegraphics[scale=0.31]{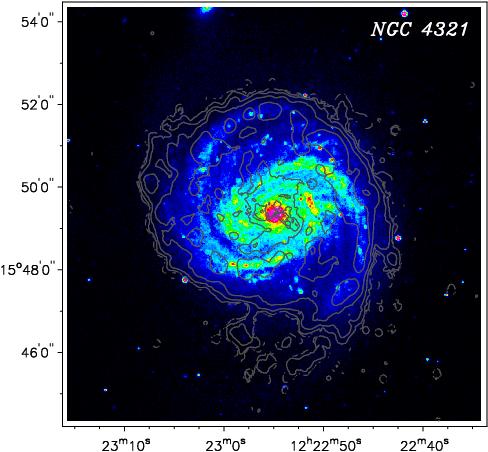}\\
\end{minipage} 
\end{figure}
\newpage
\begin{figure}[ht]
\begin{minipage}[c][21cm][t]{5.4cm}
\includegraphics[scale=0.31]{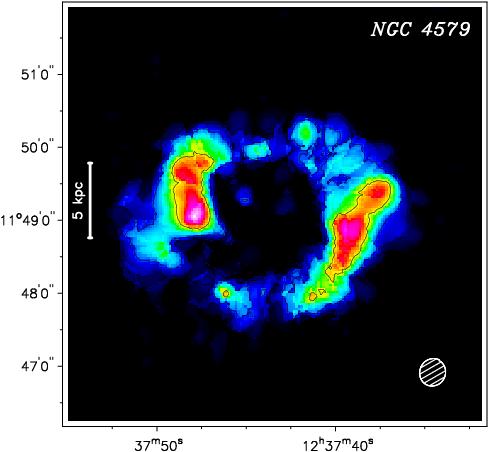}\\
\includegraphics[scale=0.31]{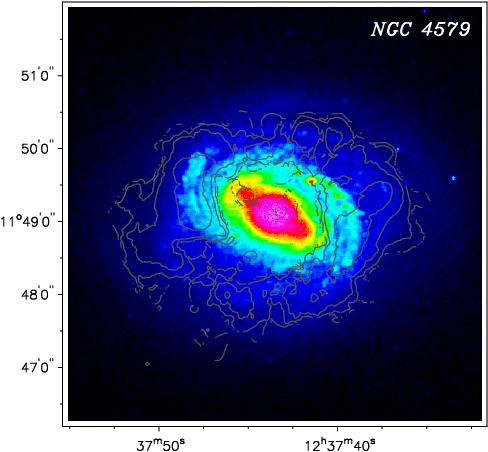}\\
\includegraphics[scale=0.31]{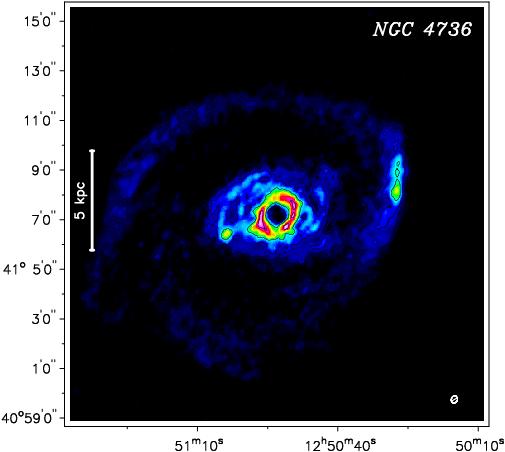}\\
\includegraphics[scale=0.31]{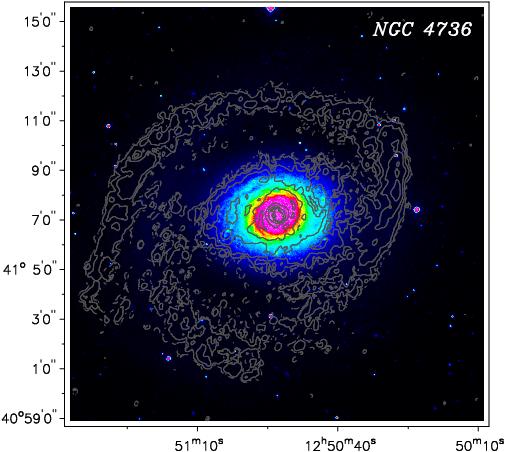}\\
\end{minipage} 
\begin{minipage}[c][21cm][t]{5.4cm}
\includegraphics[scale=0.31]{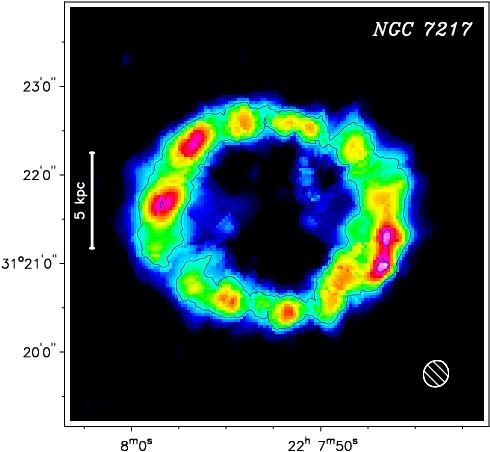}\\
\includegraphics[scale=0.31]{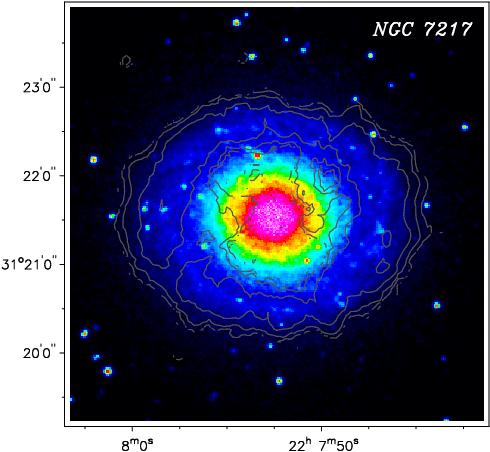}\\

\end{minipage} 
\begin{minipage}[c][21cm][t]{5.4cm}
\includegraphics[scale=0.31]{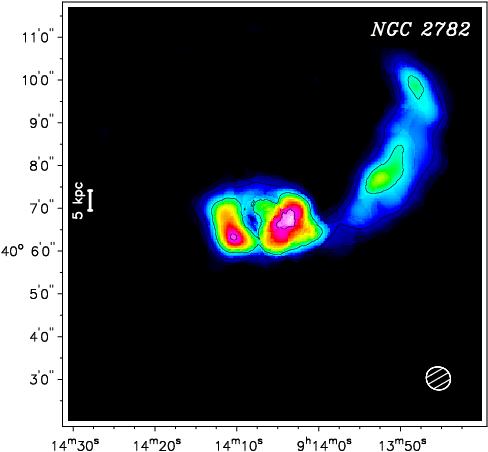}\\
\includegraphics[scale=0.31]{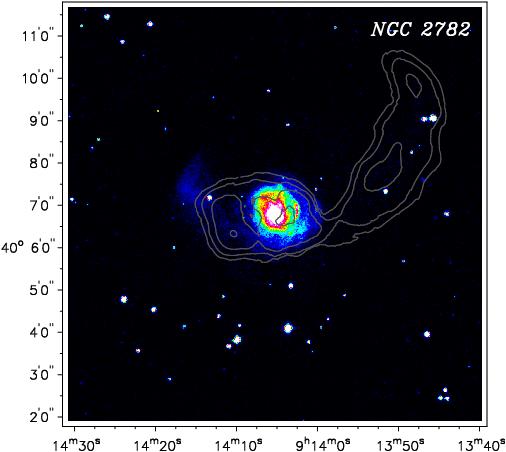}\\
\includegraphics[scale=0.31]{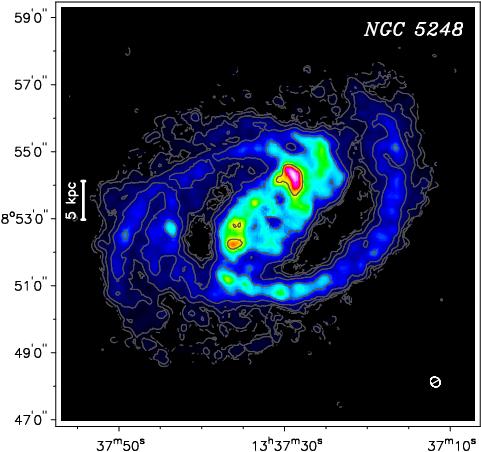}\\
\includegraphics[scale=0.31]{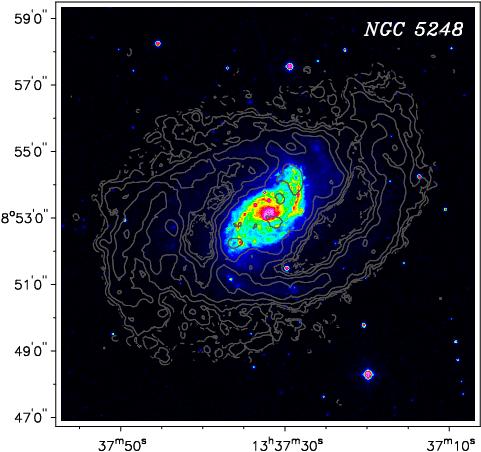}\\
\end{minipage} 
\caption{\footnotesize
HI intensity maps (upper panel) and the optical emission from DSS2-blue (lower panel with HI in contours) of our HI-NUGA sample sorted by AGN type (Seyfert, LINER, HII). The contours are in $n$ steps of  $5 \cdot 2^n \cdot 10^{19}$ cm$^{-2}$ of the HI column density (with $n=1,2,3,...$). 
\label{fig_mom0}}
\end{figure}

\clearpage

\begin{figure}[ht]
\begin{minipage}[c][21cm][t]{5.4cm}
\includegraphics[scale=0.31]{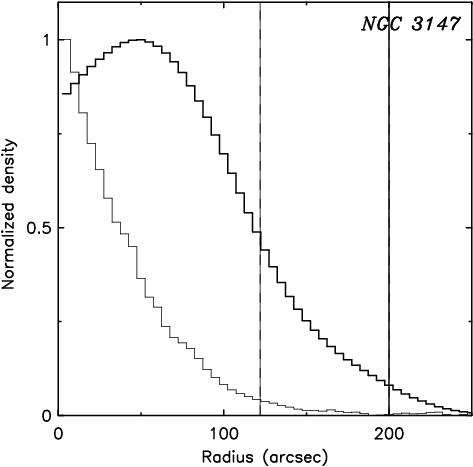}\\
\includegraphics[scale=0.31]{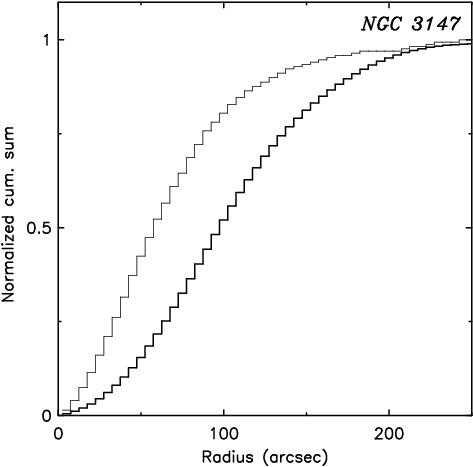}\\
\includegraphics[scale=0.31]{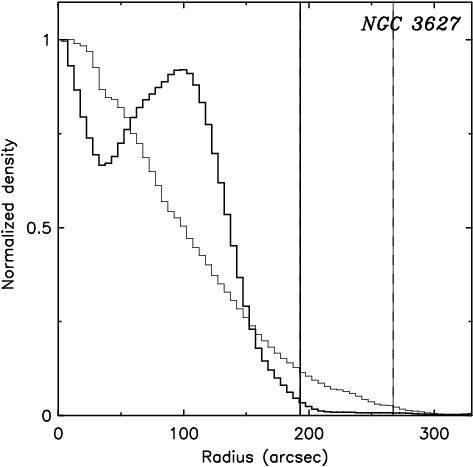}\\
\includegraphics[scale=0.31]{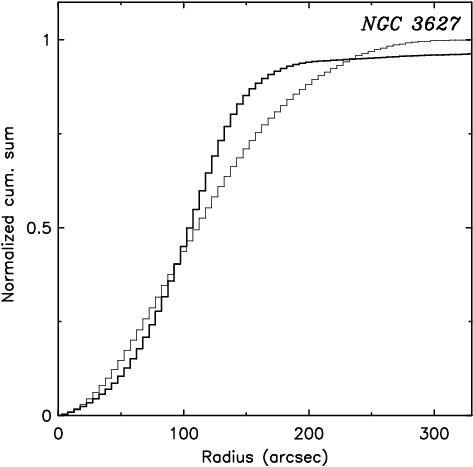}\\
\end{minipage} 
\begin{minipage}[c][21cm][t]{5.4cm}
\includegraphics[scale=0.31]{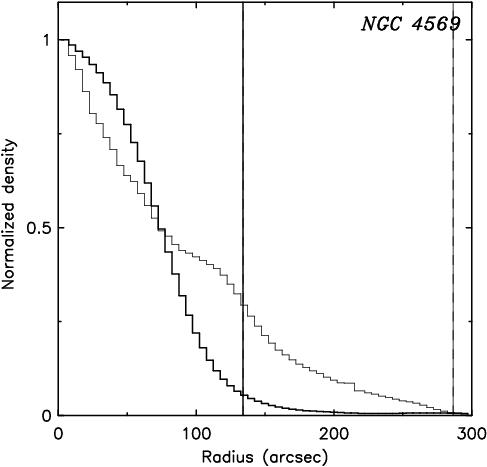}\\
\includegraphics[scale=0.31]{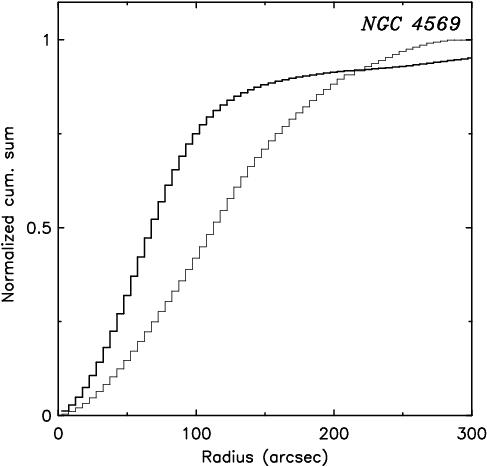}\\
\includegraphics[scale=0.31]{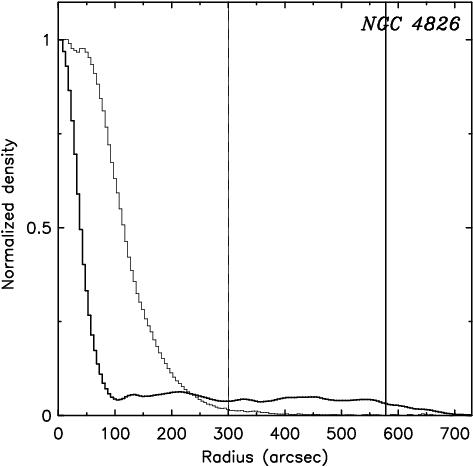}\\
\includegraphics[scale=0.31]{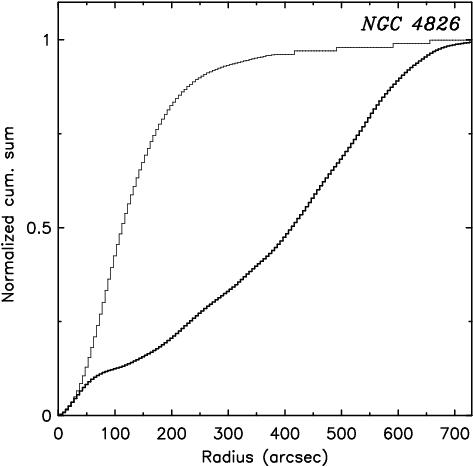}\\
\end{minipage} 
\begin{minipage}[c][21cm][t]{5.4cm}
\includegraphics[scale=0.31]{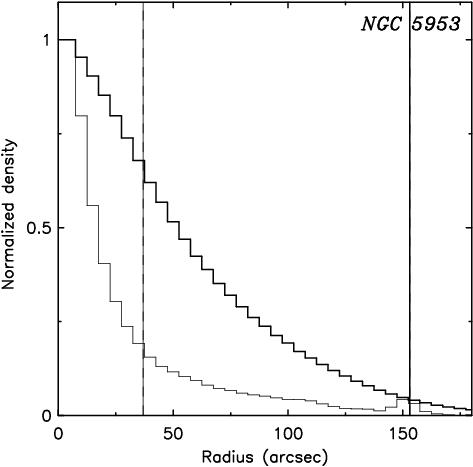}\\
\includegraphics[scale=0.31]{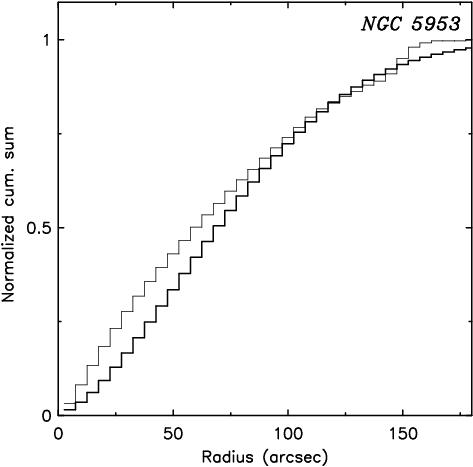}\\
\includegraphics[scale=0.31]{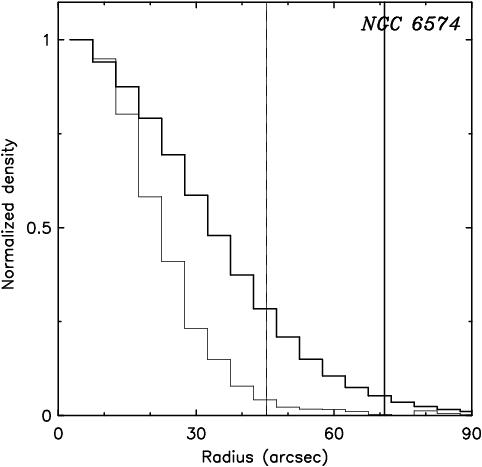}\\
\includegraphics[scale=0.31]{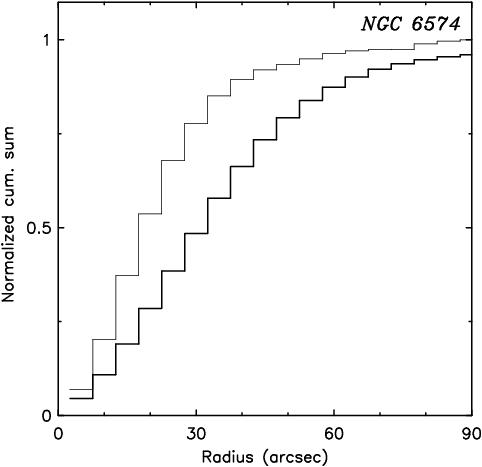}\\
\end{minipage} 
\end{figure}
\newpage
\begin{figure}[ht]
\begin{minipage}[c][21cm][t]{5.4cm}
\includegraphics[scale=0.31]{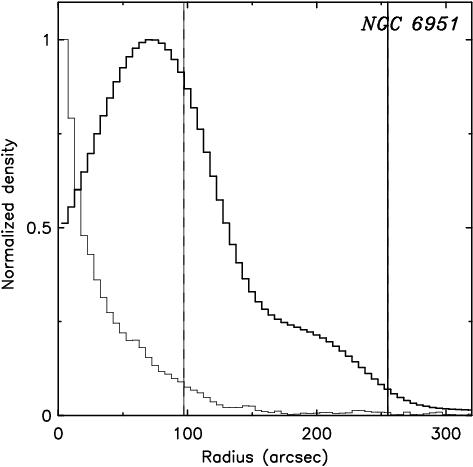}\\
\includegraphics[scale=0.31]{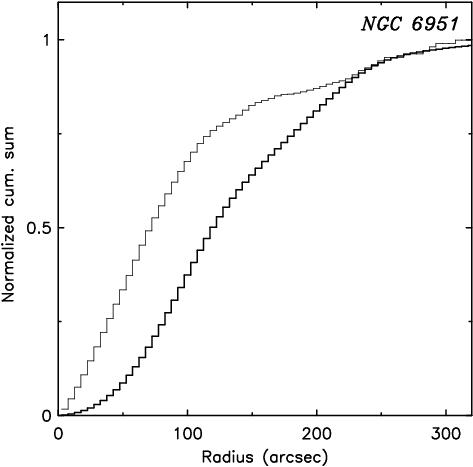}\\

\end{minipage} 
\begin{minipage}[c][21cm][t]{5.4cm}
\includegraphics[scale=0.31]{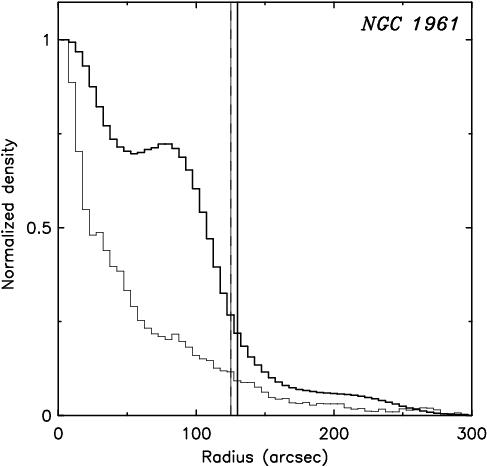}\\
\includegraphics[scale=0.31]{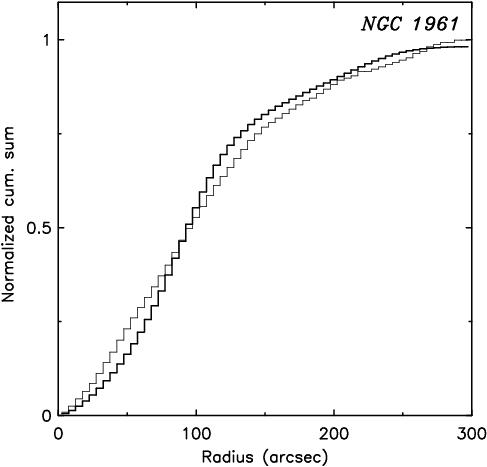}\\
\includegraphics[scale=0.31]{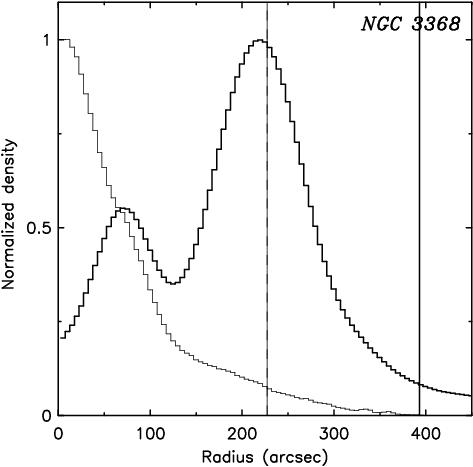}\\
\includegraphics[scale=0.31]{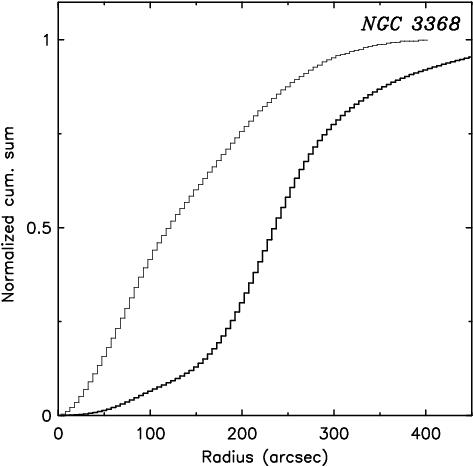}\\
\end{minipage} 
\begin{minipage}[c][21cm][t]{5.4cm}
\includegraphics[scale=0.31]{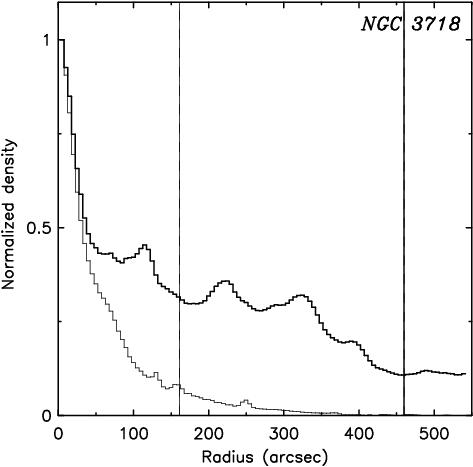}\\
\includegraphics[scale=0.31]{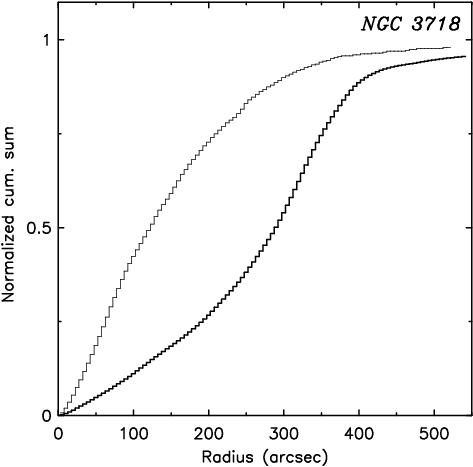}\\
\includegraphics[scale=0.31]{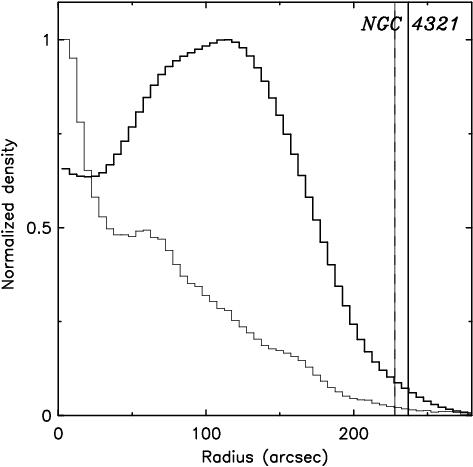}\\
\includegraphics[scale=0.31]{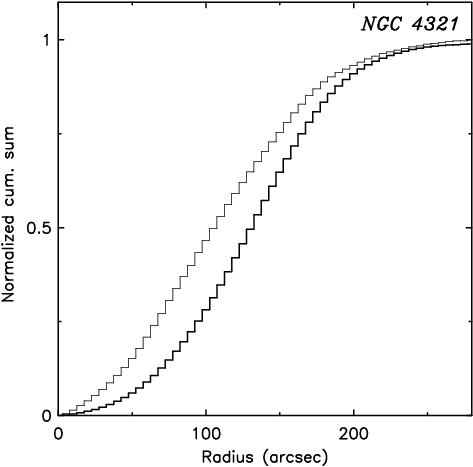}\\
\end{minipage} 
\end{figure}
\newpage
\begin{figure}[ht]
\begin{minipage}[c][21cm][t]{5.4cm}
\includegraphics[scale=0.31]{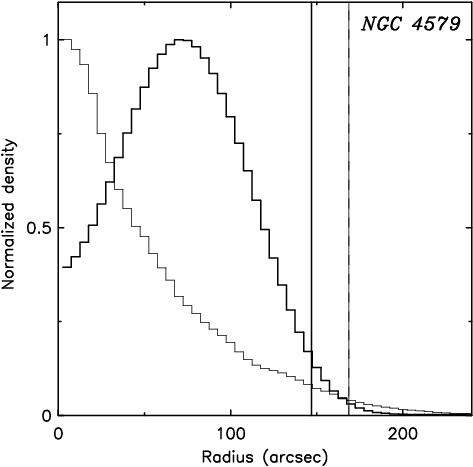}\\
\includegraphics[scale=0.31]{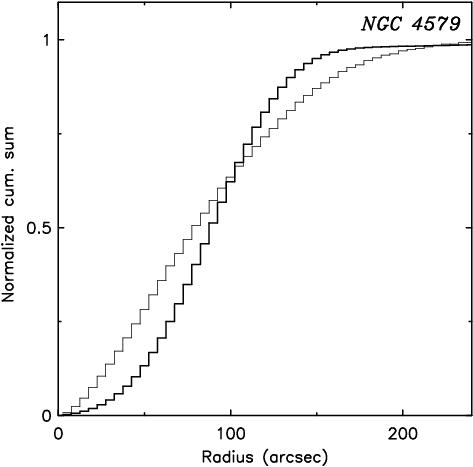}\\
\includegraphics[scale=0.31]{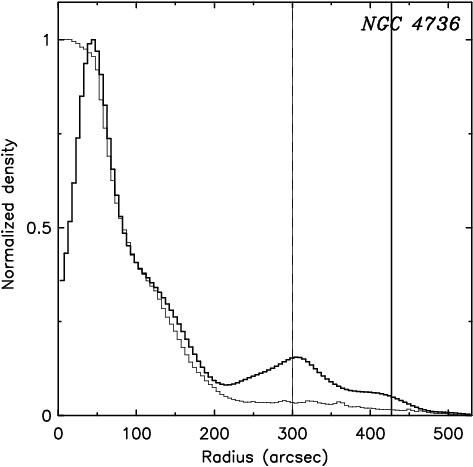}\\
\includegraphics[scale=0.31]{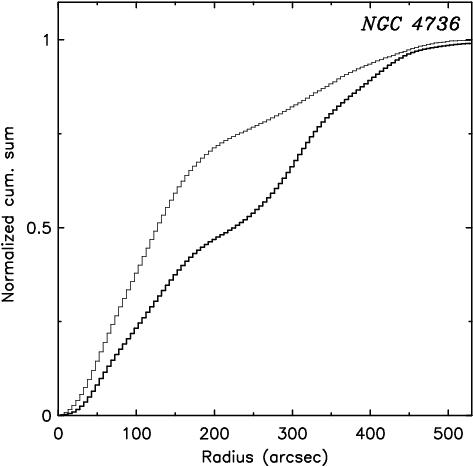}\\
\end{minipage} 
\begin{minipage}[c][21cm][t]{5.4cm}
\includegraphics[scale=0.31]{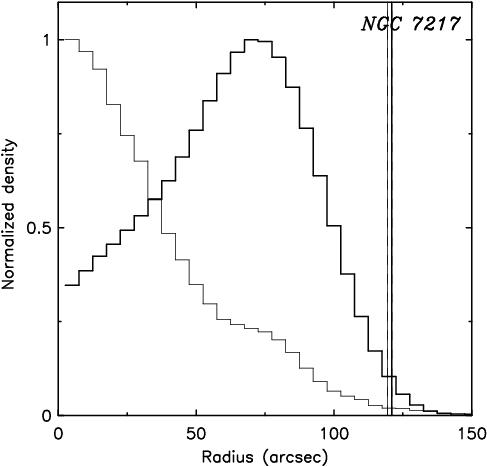}\\
\includegraphics[scale=0.31]{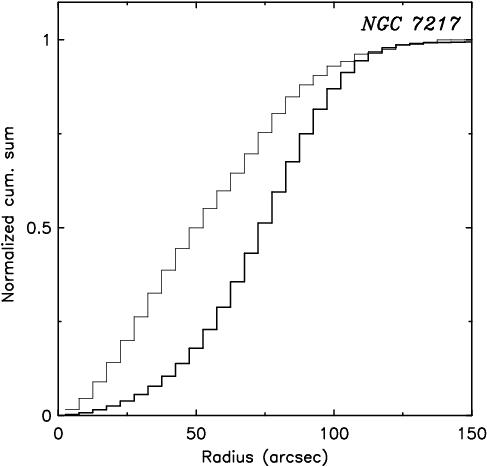}\\

\end{minipage} 
\begin{minipage}[c][21cm][t]{5.4cm}
\includegraphics[scale=0.31]{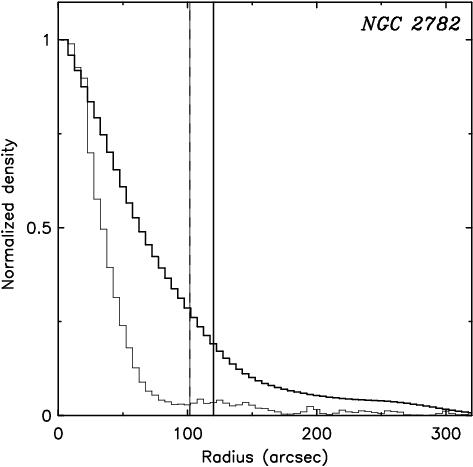}\\
\includegraphics[scale=0.31]{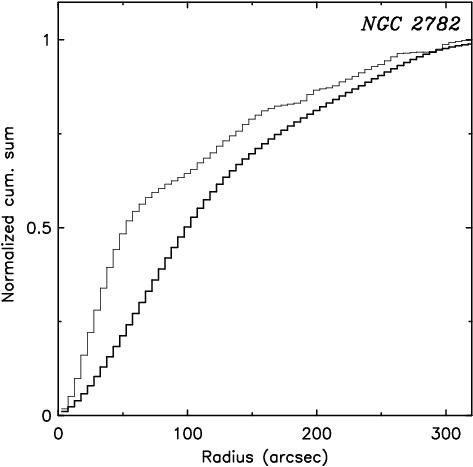}\\
\includegraphics[scale=0.31]{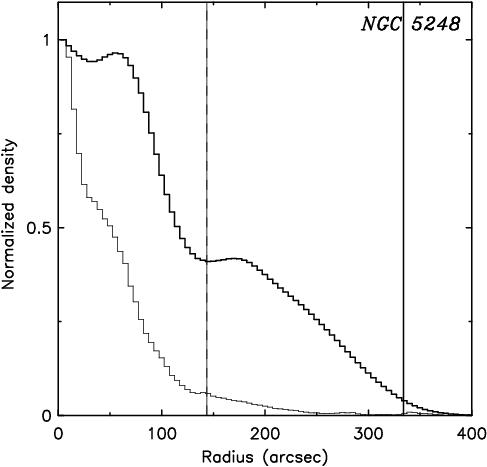}\\
\includegraphics[scale=0.31]{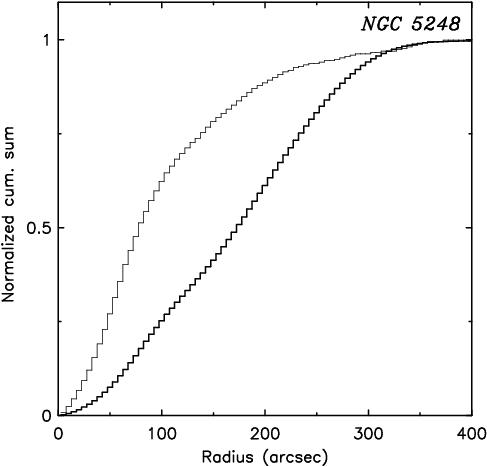}\\
\end{minipage} 
\caption{\footnotesize
Comparison between the optical (thin line) and HI (thick line) radial distribution. The mean radial HI and optical profile are presented in the top panel and the cumulative profiles of the HI and optical emission in the bottom panel for each galaxy. The radius of the HI disk $R_{HI}$ (thick line) and the radius of the stellar disk $R_{B25}$ (thin line) are indicated. The values of $R_{B25}$ are taken from the parameter log(d$_{B25}$) found in the literature (Hyperleda) and the radius $R_{HI}$ for the HI disk corresponds to a measured column density of $5.0\times 10^{19}$ cm$^{-2}$.
\label{fig_rad}}
\end{figure}

\clearpage

\begin{figure}[ht]
\begin{center}
\includegraphics[scale=0.5]{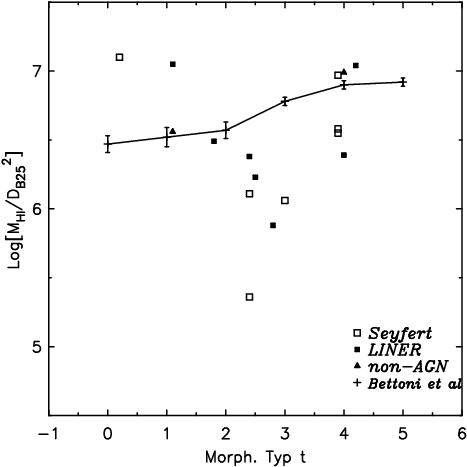}
\caption{\footnotesize
Comparison of the relative gas content versus Hubble type (Morphology type t) between our sample (Sy, LINER, non-AGN) and a larger sample of 1916 galaxies by Bettoni et al (\citeyear{Bet03}; the average values are plotted and connected with a solid line). The relative gas content is given as the ratio of the atomic gas mass to the square of the diameter in kpc at the isophote of 25 mag arcsec$^{−2}$ $M_{HI}/D_{B25}^2$.
\label{fig_Hubble}}
\end{center}
\end{figure}

\newpage

\begin{figure}[ht]
\begin{minipage}[c][21cm][t]{5.4cm}
\includegraphics[scale=0.38]{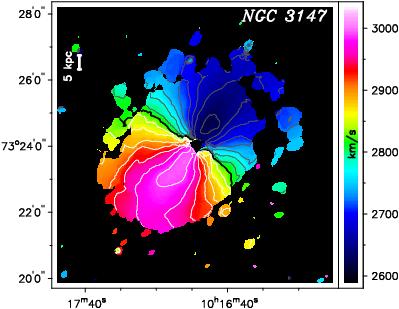}\\
\includegraphics[scale=0.38]{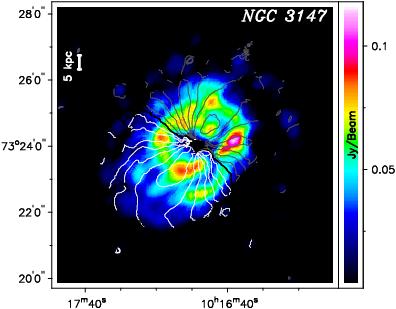}\\
\includegraphics[scale=0.38]{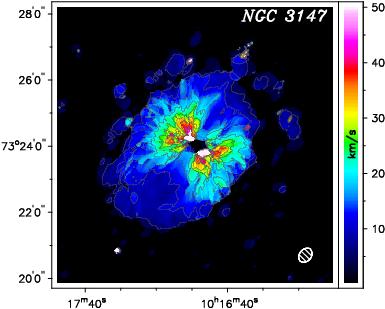}\\
\includegraphics[scale=0.38]{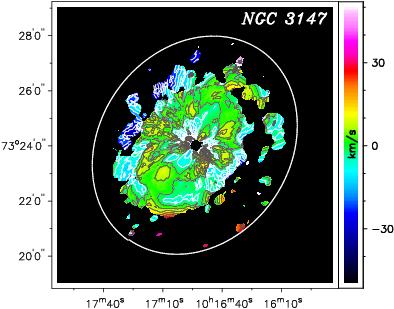}\\
\end{minipage} 
\begin{minipage}[c][21cm][t]{5.4cm}
\includegraphics[scale=0.38]{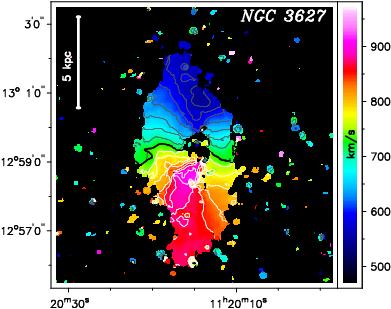}\\
\includegraphics[scale=0.38]{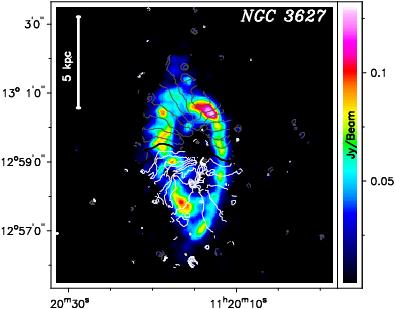}\\
\includegraphics[scale=0.38]{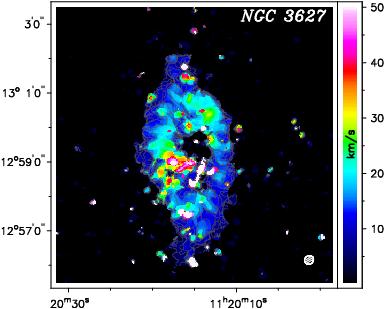}\\
\includegraphics[scale=0.38]{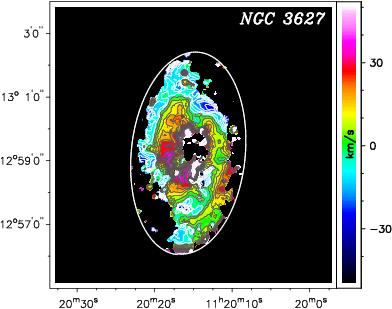}\\
\end{minipage} 
\begin{minipage}[c][21cm][t]{5.4cm}
\includegraphics[scale=0.38]{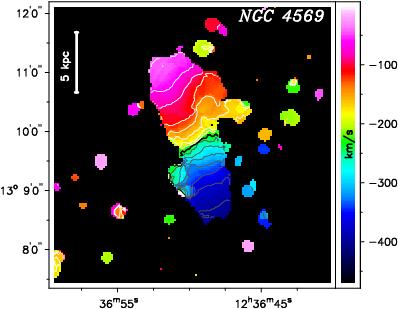}\\
\includegraphics[scale=0.38]{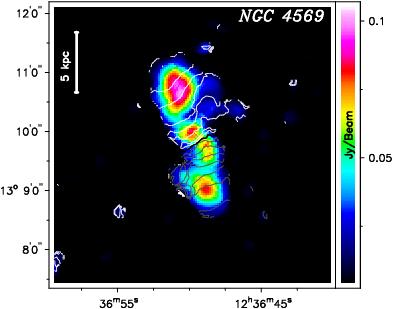}\\
\includegraphics[scale=0.38]{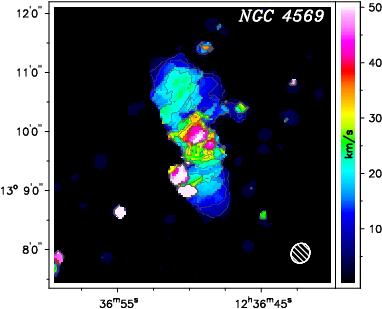}\\
\includegraphics[scale=0.38]{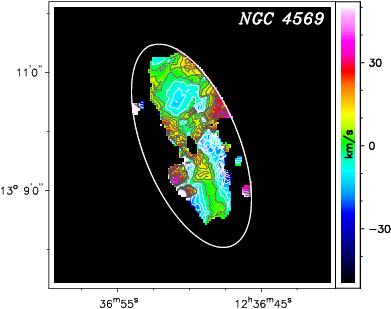}\\
\end{minipage} 
\end{figure}
\newpage
\begin{figure}[ht]
\begin{minipage}[c][21cm][t]{5.4cm}
\includegraphics[scale=0.38]{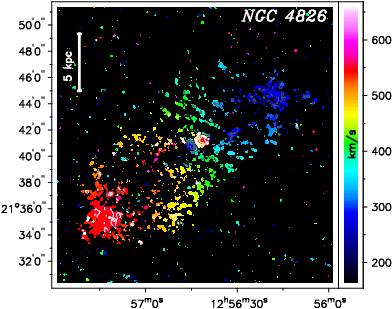}\\
\includegraphics[scale=0.38]{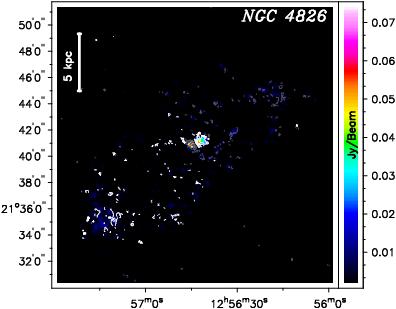}\\
\includegraphics[scale=0.38]{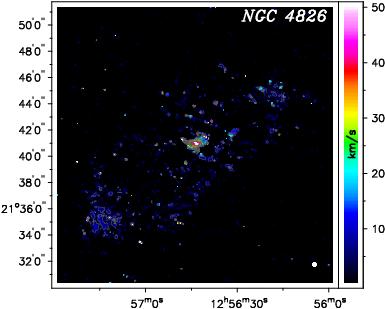}\\
\includegraphics[scale=0.38]{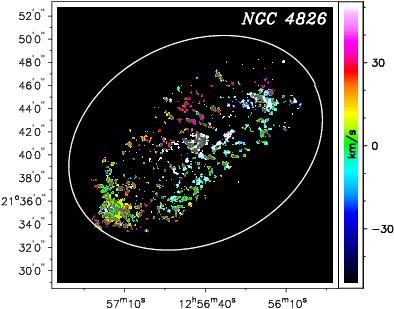}\\
\end{minipage} 
\begin{minipage}[c][21cm][t]{5.4cm}
\includegraphics[scale=0.38]{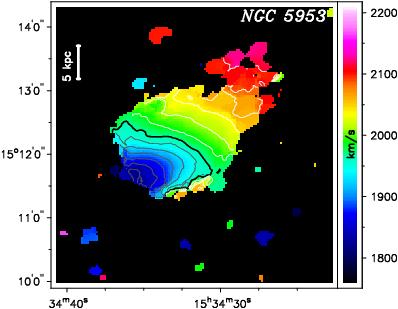}\\
\includegraphics[scale=0.38]{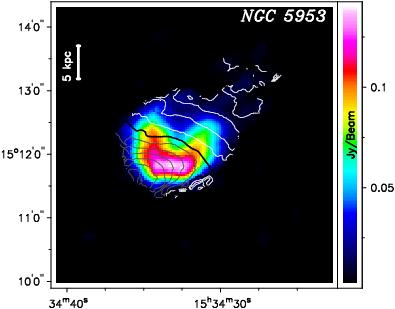}\\
\includegraphics[scale=0.38]{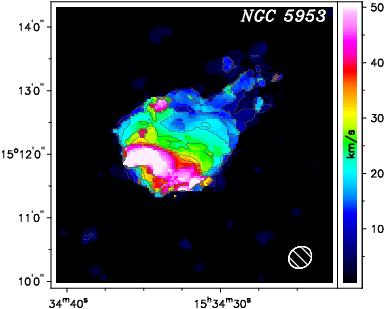}\\
\includegraphics[scale=0.38]{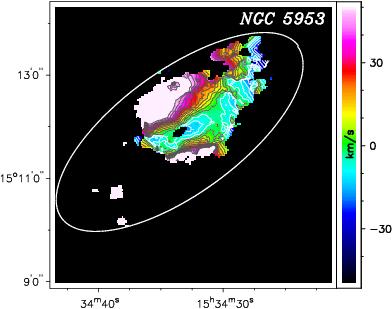}\\
\end{minipage} 
\begin{minipage}[c][21cm][t]{5.4cm}
\includegraphics[scale=0.38]{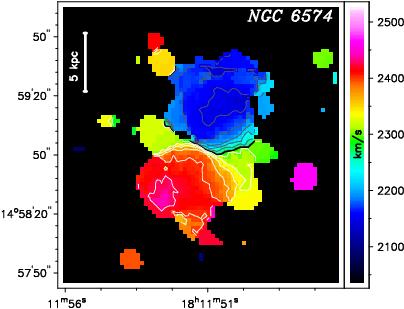}\\
\includegraphics[scale=0.38]{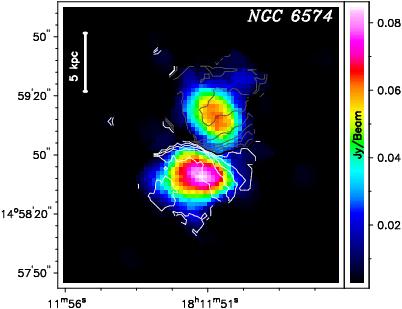}\\
\includegraphics[scale=0.38]{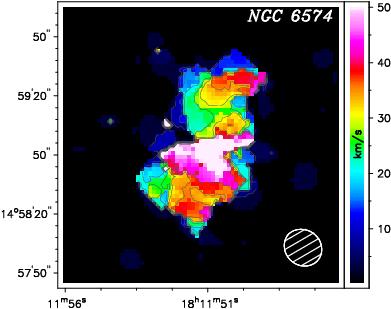}\\
\includegraphics[scale=0.38]{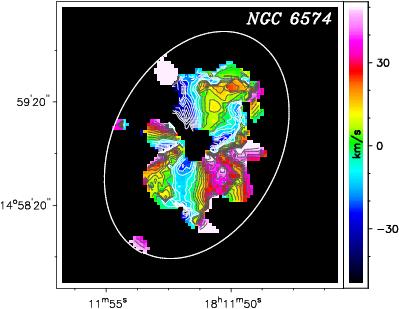}\\
\end{minipage} 
\end{figure}
\newpage
\begin{figure}[ht]
\begin{minipage}[c][21cm][t]{5.4cm}
\includegraphics[scale=0.38]{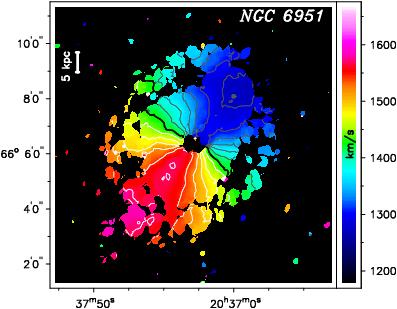}\\
\includegraphics[scale=0.38]{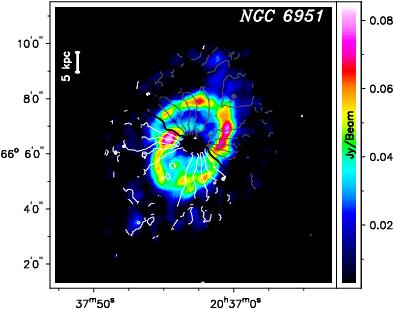}\\
\includegraphics[scale=0.38]{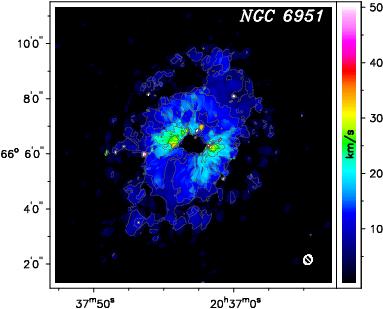}\\
\includegraphics[scale=0.38]{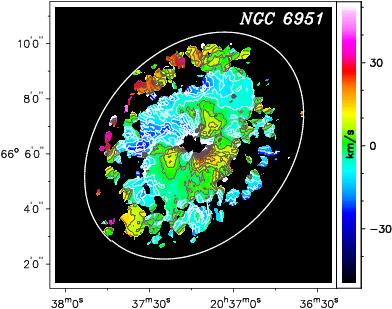}\\
\end{minipage} 
\begin{minipage}[c][21cm][t]{5.4cm}

\end{minipage} 
\begin{minipage}[c][21cm][t]{5.4cm}
\includegraphics[scale=0.38]{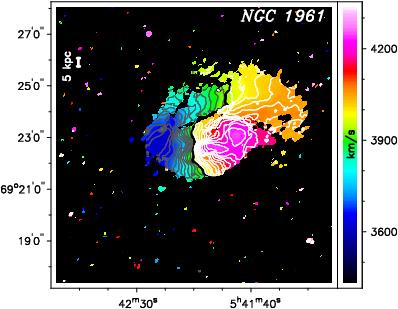}\\
\includegraphics[scale=0.38]{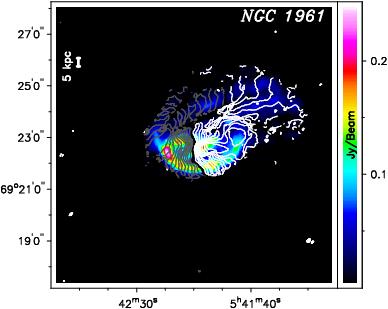}\\
\includegraphics[scale=0.38]{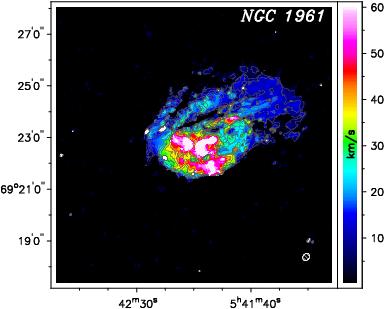}\\
\includegraphics[scale=0.38]{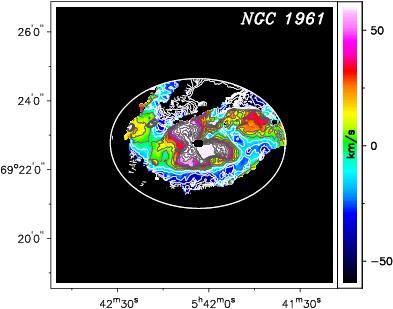}\\
\end{minipage} 
\end{figure}
\newpage
\begin{figure}[ht]
\begin{minipage}[c][21cm][t]{5.4cm}
\includegraphics[scale=0.38]{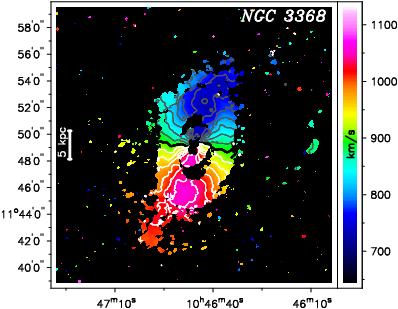}\\
\includegraphics[scale=0.38]{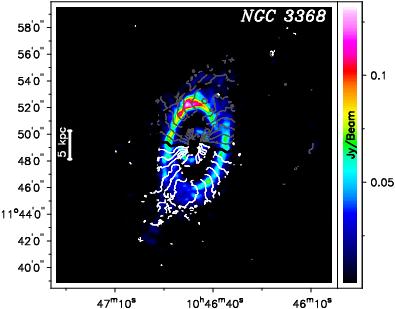}\\
\includegraphics[scale=0.38]{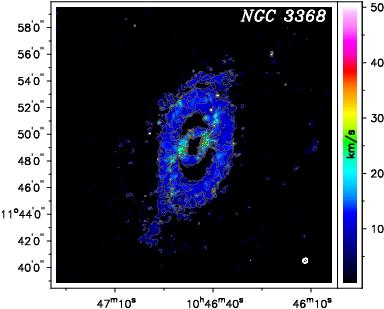}\\
\includegraphics[scale=0.38]{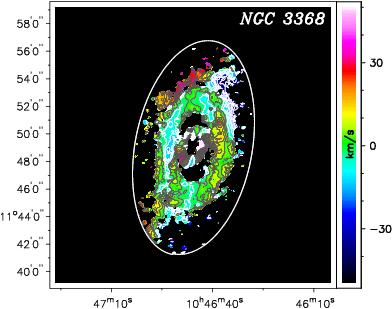}\\
\end{minipage} 
\begin{minipage}[c][21cm][t]{5.4cm}
\includegraphics[scale=0.38]{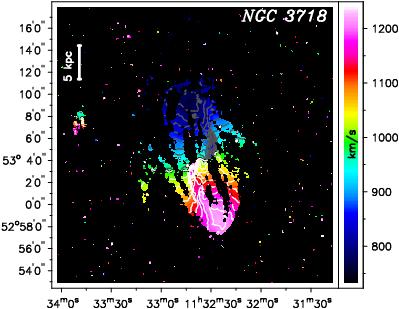}\\
\includegraphics[scale=0.38]{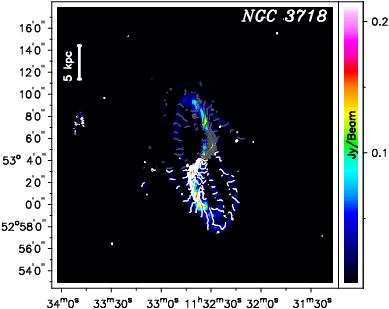}\\
\includegraphics[scale=0.38]{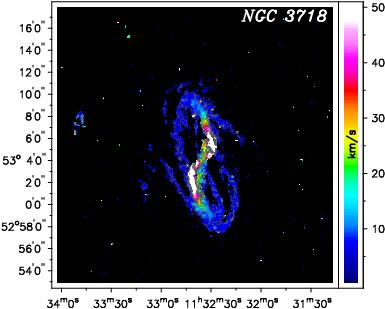}\\
\includegraphics[scale=0.38]{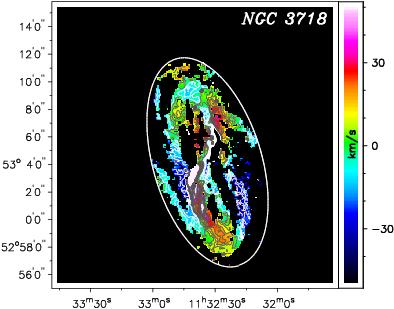}\\
\end{minipage} 
\begin{minipage}[c][21cm][t]{5.4cm}
\includegraphics[scale=0.38]{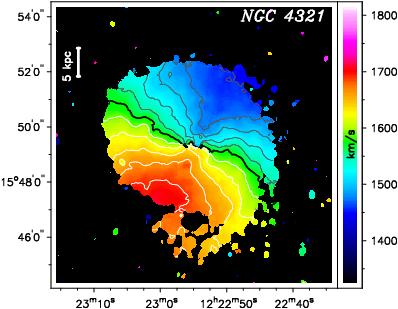}\\
\includegraphics[scale=0.38]{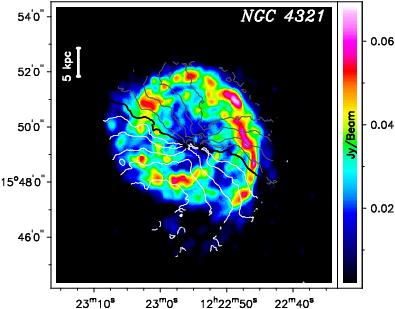}\\
\includegraphics[scale=0.38]{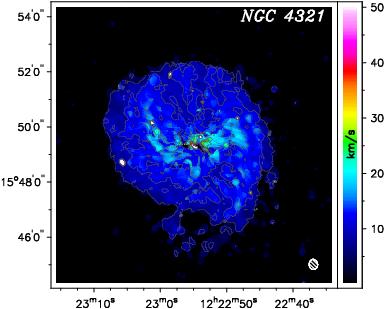}\\
\includegraphics[scale=0.38]{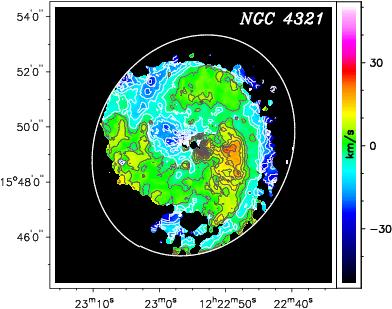}\\
\end{minipage} 
\end{figure}
\newpage
\begin{figure}[ht]
\begin{minipage}[c][21cm][t]{5.4cm}
\includegraphics[scale=0.38]{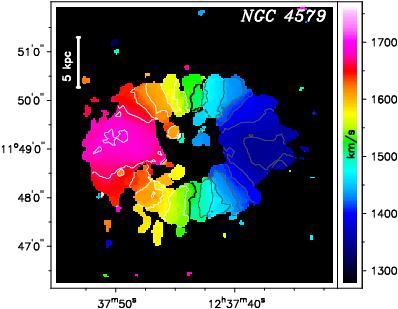}\\
\includegraphics[scale=0.38]{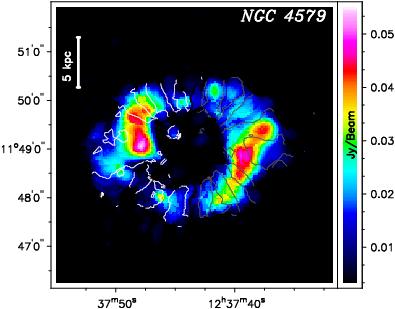}\\
\includegraphics[scale=0.38]{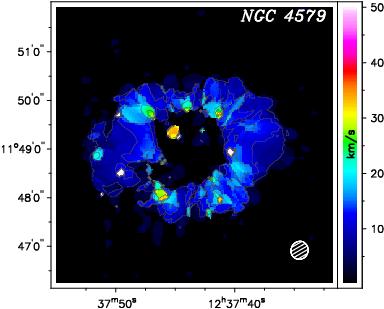}\\
\includegraphics[scale=0.38]{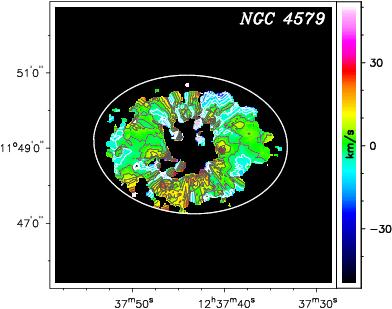}\\
\end{minipage} 
\begin{minipage}[c][21cm][t]{5.4cm}
\includegraphics[scale=0.38]{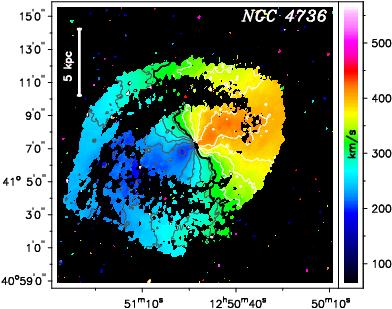}\\
\includegraphics[scale=0.38]{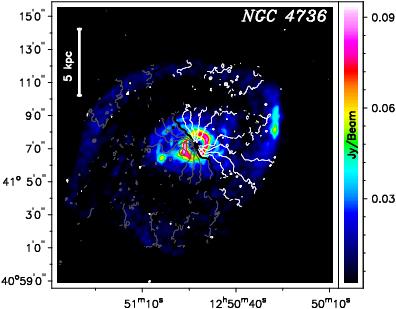}\\
\includegraphics[scale=0.38]{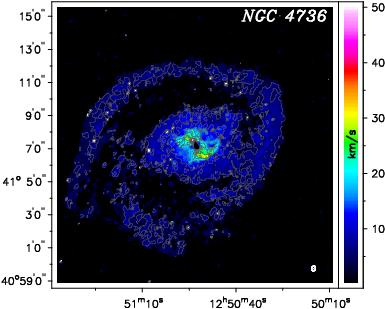}\\
\includegraphics[scale=0.38]{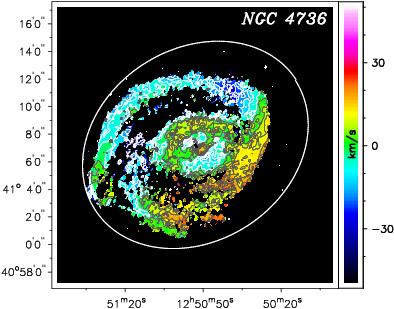}\\
\end{minipage} 
\begin{minipage}[c][21cm][t]{5.4cm}
\includegraphics[scale=0.38]{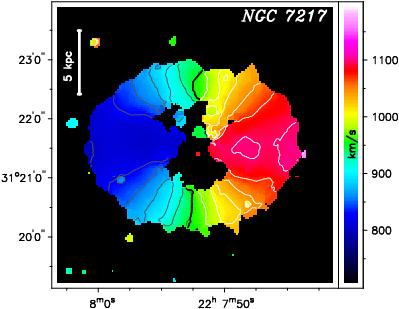}\\
\includegraphics[scale=0.38]{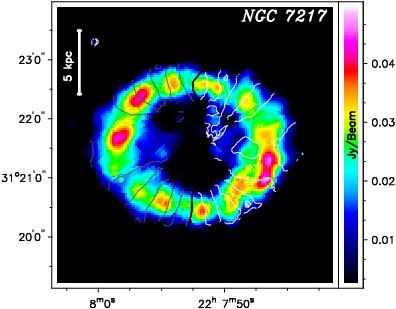}\\
\includegraphics[scale=0.38]{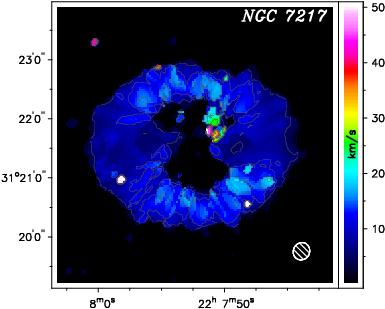}\\
\includegraphics[scale=0.38]{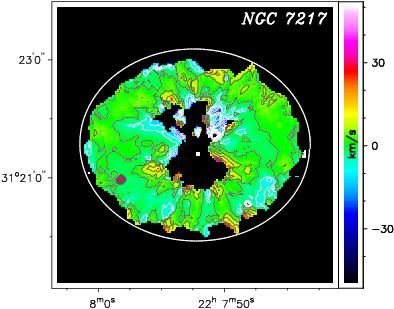}\\
\end{minipage} 
\end{figure}
\newpage
\begin{figure}[ht]
\begin{minipage}[c][21cm][t]{5.4cm}
\includegraphics[scale=0.38]{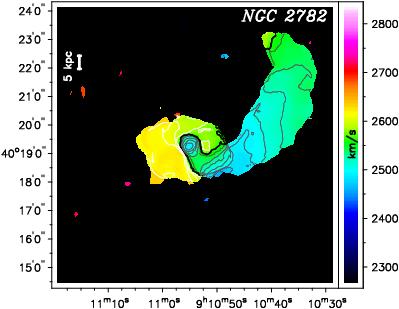}\\
\includegraphics[scale=0.38]{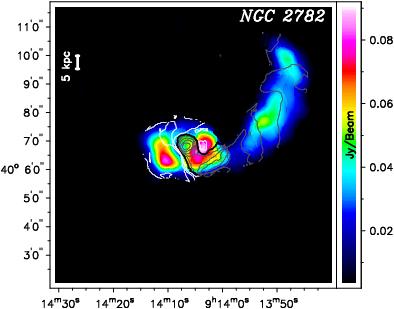}\\
\includegraphics[scale=0.38]{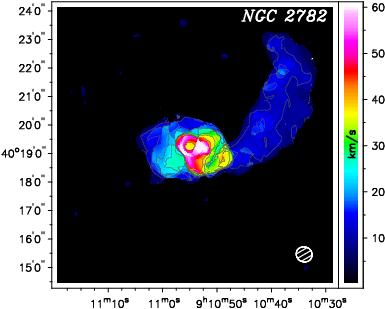}\\
\includegraphics[scale=0.38]{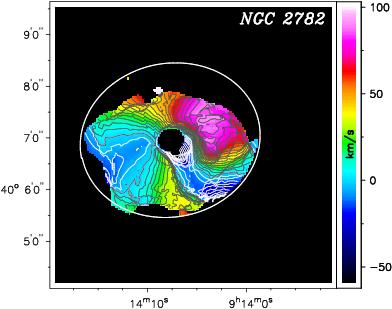}\\
\end{minipage} 
\begin{minipage}[c][21cm][t]{5.4cm}
\includegraphics[scale=0.38]{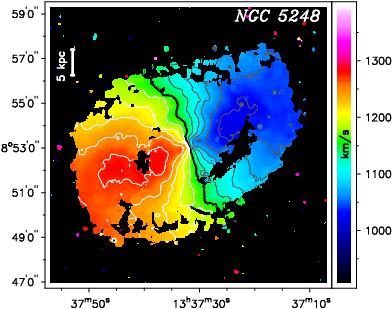}\\
\includegraphics[scale=0.38]{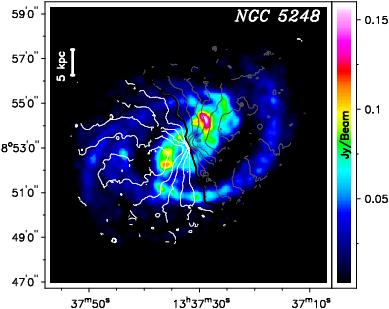}\\
\includegraphics[scale=0.38]{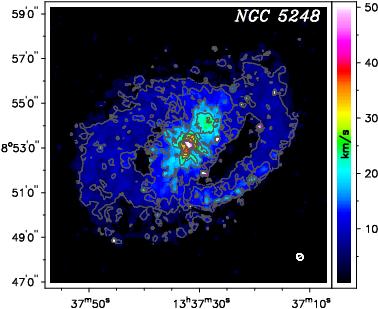}\\
\includegraphics[scale=0.38]{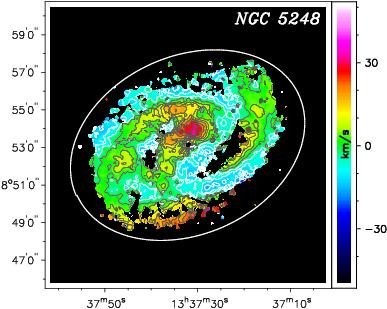}\\
\end{minipage} 
\begin{minipage}[c][21cm][t]{5.4cm}

\end{minipage} 
\vspace{-2.5cm}
\caption{\footnotesize
Overview of all HI kinematic maps for each galaxy. The intensity-weighted HI mean velocity field, the comparison of HI intensity map and velocity field, the velocity dispersion map, and the residual velocity field are presented (from the top to the bottom).
The contours are set in an interval of 25 km s$^{-1}$ for the isovelocity lines (first and second panel from the top), the central line (thick dark line) represents the systemic velocity and white (grey) lines the approaching (receding) velocities. For the dispersion and residual field contours are set in 5 km s$^{-1}$ intervals. The residual velocity field was created by subtracting a model velocity field, derived from a smooth fit to the rotation curve, from the observed velocity field (see text for details). The white ellipse in the residual field (bottom panel) indicates the maximum radius of the model velocity field.
\label{fig_kin}}
\end{figure}

\clearpage

\begin{figure}[ht]
\begin{center}
\resizebox{1\hsize}{!}{\fbox{\includegraphics{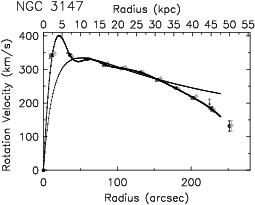}
                       \includegraphics{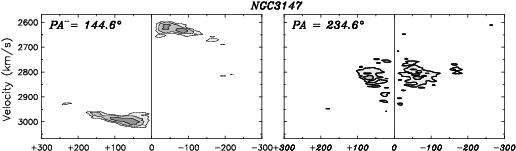}}}\\
\resizebox{1\hsize}{!}{\fbox{\includegraphics{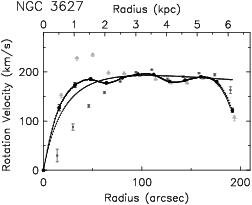}
                       \includegraphics{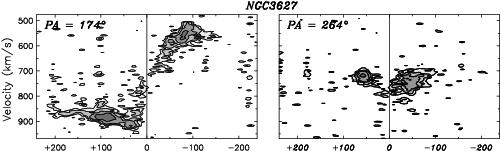}}}\\
\resizebox{1\hsize}{!}{\fbox{\includegraphics{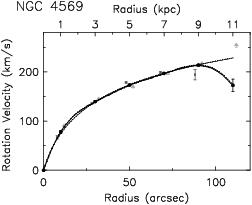}
                       \includegraphics{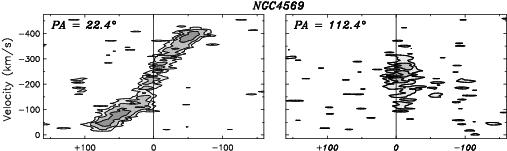}}}\\ 
\resizebox{1\hsize}{!}{\fbox{\includegraphics{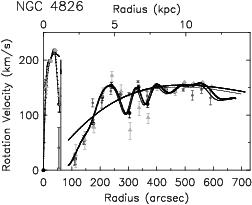}
                       \includegraphics{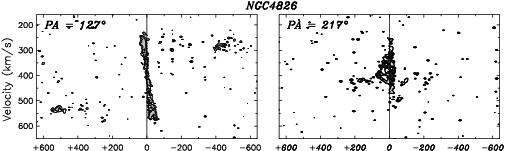}}}\\ 

\end{center}
\end{figure}
\begin{figure}[ht]
\begin{center}
\resizebox{1\hsize}{!}{\fbox{\includegraphics{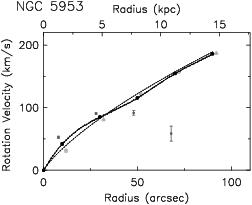}
                       \includegraphics{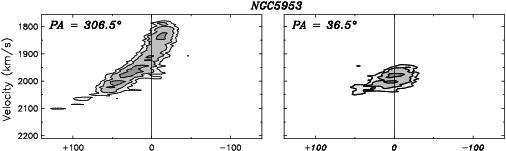}}}\\ 
\resizebox{1\hsize}{!}{\fbox{\includegraphics{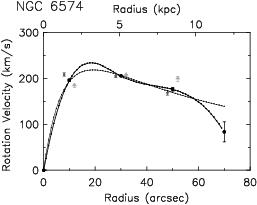}
                       \includegraphics{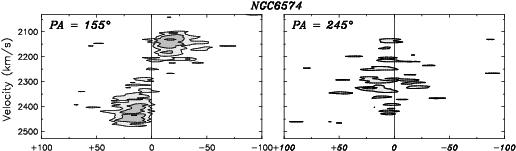}}}\\ 
\resizebox{1\hsize}{!}{\fbox{\includegraphics{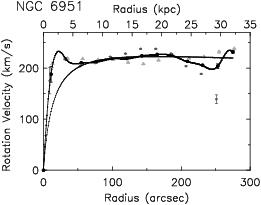}
                       \includegraphics{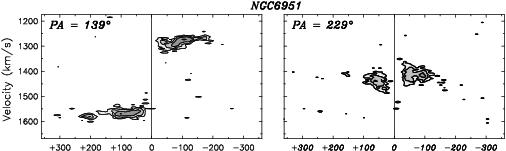}}}\\ 
\end{center}
\end{figure}
\begin{figure}[ht]
\begin{center}
\resizebox{1\hsize}{!}{\fbox{\includegraphics{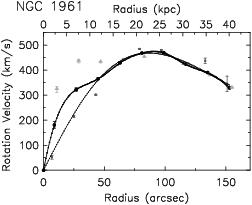}
                       \includegraphics{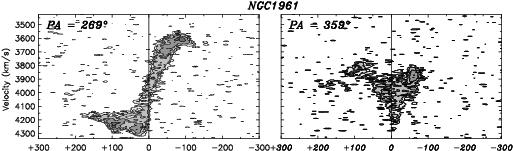}}}\\ 
\resizebox{1\hsize}{!}{\fbox{\includegraphics{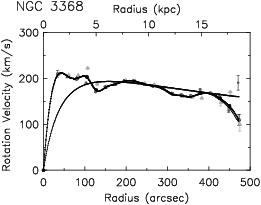}
                       \includegraphics{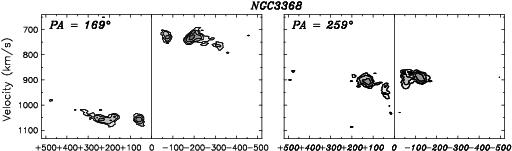}}}\\ 
\resizebox{1\hsize}{!}{\fbox{\includegraphics{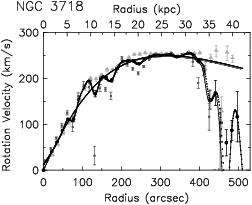}
                       \includegraphics{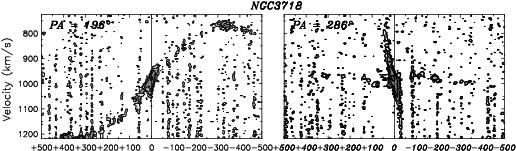}}}\\ 
\resizebox{1\hsize}{!}{\fbox{\includegraphics{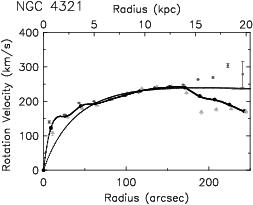}
                       \includegraphics{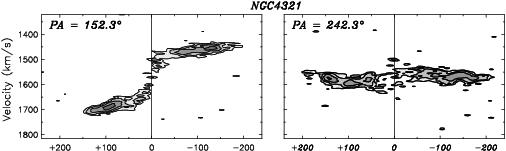}}}\\

\end{center}
\end{figure}
\begin{figure}[ht]
\begin{center}
\resizebox{1\hsize}{!}{\fbox{\includegraphics{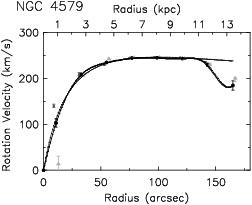}
                       \includegraphics{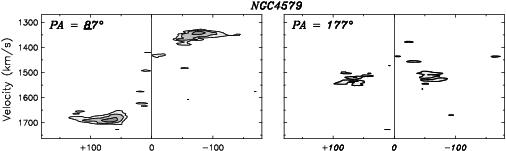}}}\\ 
\resizebox{1\hsize}{!}{\fbox{\includegraphics{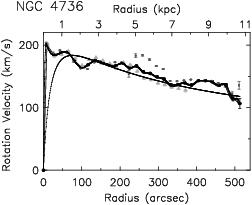}
                       \includegraphics{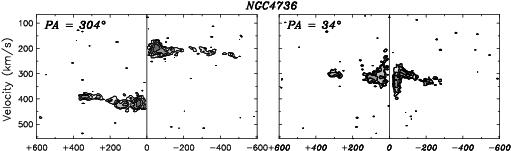}}}\\ 
\resizebox{1\hsize}{!}{\fbox{\includegraphics{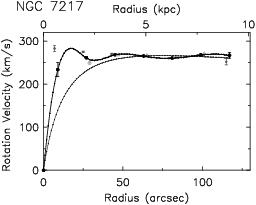}
                       \includegraphics{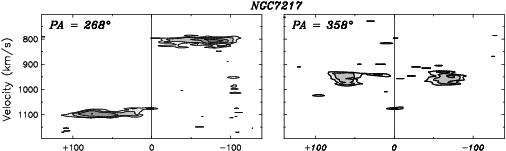}}}\\ 
\end{center}
\end{figure}
\begin{figure}[ht]
\begin{center}
\resizebox{1\hsize}{!}{\fbox{\includegraphics{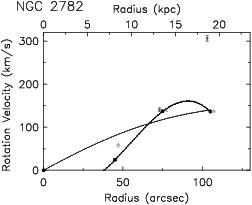}
                       \includegraphics{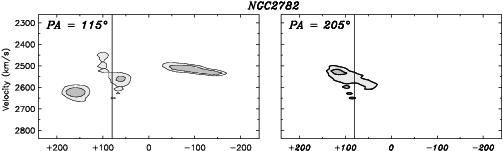}}}\\ 
\resizebox{1\hsize}{!}{\fbox{\includegraphics{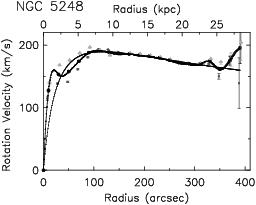}
                       \includegraphics{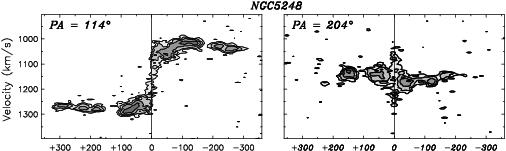}}}\\ 
\caption{\footnotesize
Presentation of the rotation curve and pv-diagrams for each galaxy. The rotation velocities have been plotted in the left panel for a) the whole disk (black dots), b) the approaching side of the disk only (dark grey dots), and c) the receding side of the disk only (light grey dots). A Brandt curve \citep{Brandt} is fitted (thin black line) as well as a spline interpolation (thick black line) to the derived rotation velocities of the whole disk. The center and right panel show the pv diagram of the major and minor axis, respectively. The contours are in $n$ steps of the column density at $1.8 \cdot 2^n \cdot 10^{19}$ cm$^{-2}$ ($n=1,2,3,...$). Note that the values for NGC~2782 are likely incorrect, as there is a short tidal tail to the east which is barely resolved at our resolution \citep[see][]{Smi94}.
\label{fig_rot}}
\end{center}
\end{figure}

\clearpage

\begin{table}
\begin{footnotesize}
\caption{Sample Overview}
\begin{tabular}{lcclccrcll}
\tableline\tableline
Name & RA & DEC & Hubble & Bar & AGN& AGN & $v_{hel}$ & Dist & Res.\\
       & (J2000) & (J2000) & Type & & HFS & Kewley & [km/s]&[Mpc]&[kpc]\\ 
\tableline 
NGC~3147& 10 16 53.65 & +73 24 02.7 & S(rs)bc  &   & S2 & S & 2820 & 40.9 & 5.0\\
NGC~3627& 11 20 15.03 & +12 59 29.6 & SB(s)b  & x & L/S2 & S &  727 & 6.6 & 0.5\\
NGC~4569& 12 36 49.80 & +13 09 46.3 & SB(rs)ab & x & T2 & S & -235 & 16.8 & 1.6\\
NGC~4826& 12 56 43.69 & +21 40 57.5 & (R)S(rs)ab  &   & T2 & S &  408 & 4.1 & 0.3\\
NGC~5953& 15 34 32.39 & +15 11 37.7 & SO-a &   & S2 & S & 1965 & 36 & 3.3\\
NGC~6574& 18 11 51.23 & +14 58 54.4 & SB(rs)bc & x & S & S & 2282 & 38 & 3.3\\
NGC~6951& 20 37 14.09 & +66 06 20.3 & SB(rs)bc & x & S2 & S & 1424 & 24.1 & 2.4\\
\tableline
NGC~1961& 05 42 04.80 & +69 22 43.3 & SB(rs)bc & x & L2 & L & 3934 & 53.9 & 4.4\\
NGC~3368& 10 46 45.74 & +11 49 11.8 & SB(rs)ab & x & L2 & L &  897 & 8.1 & 0.8\\
NGC~3718& 11 32 34.85 & +53 04 04.5 & SB(s)a  & x & L1.9& L & 994 & 17.0 & 1.1\\
NGC~4321& 12 22 54.90 & +15 49 20.6 & SB(s)bc & x & T2 & L & 1571 & 16.8 & 1.4\\
NGC~4579& 12 37 43.52 & +11 49 05.5 & SB(rs)b  & x & S1.9/L1.9 & L & 1519 & 37 & 1.8\\
NGC~4736& 12 50 53.06 & +41 07 13.7 & (R)S(r)ab  &  x & L2 & L &  308 & 4.3 & 0.4\\
NGC~7217& 22 07 52.38 & +31 21 33.4 & (R)S(r)ab   &   & L2 & L &  952 & 16.0 & 1.4\\
\tableline
NGC~2782& 09 14 05.11 & +40 06 49.2 & SB(rs)a  & x & H & H & 2562 & 37.3 & 6.2\\
NGC~5248& 13 37 32.07 & +08 53 06.2 & (R)SB(rs)bc & x & H & H & 1153 & 15 & 1.4\\
\tableline
\end{tabular}
\tablecomments{Summary of the properties of the complete HI-NUGA sample. Listed are only parameters from LEDA, NED and the resolution of the VLA observations for all galaxies (beam).  The AGN classification listed in column (6) is taken from Ho, Fillipenko \& Sargent (1997): S - Seyfert, L - LINER, T - transition object, H - HII galaxy and NED. The galaxies are divided into separated classes as described in \S \ref{subsec:obs_sample}: Seyfert galaxies(top part), LINER galaxies (mid part) and HII/Transition objects (bottom part), additionally indicated in column (7). The velocities listed in column (8) are the assumed systemic velocities. In column (10) the spatial resolution of our HI images is listed.}
\label{table_intro}
\end{footnotesize}
\end{table}

\begin{table}
\caption{Observational Setup}
\begin{tabular}{lrrrrl}
\tableline\tableline
Name   & nat. weight & rob. weight & RMS nat. &  RMS rob. & FWZI.\\
 & Beam [$\arcsec$] & Beam [$\arcsec$] & [mJy/beam]& [mJy/beam]& [km/s]\\ 
\tableline 
NGC~3147&45.4 $\times$ 43.6&29.2 $\times$ 22.0&0.44&0.59&430\\
NGC~3627&22.5 $\times$ 19.2&16.8 $\times$ 15.2&0.34&0.68&431\\
NGC~4569&41.8 $\times$ 36.0&21.2 $\times$ 19.2&0.48&0.63&396\\
NGC~4826&25.6 $\times$ 21.6&16.3 $\times$ 14.8&0.32&0.28&372\\
NGC~5953&42.1 $\times$ 32.5&22.9 $\times$ 18.3&0.31&0.41&349\\
NGC~6574&30.5 $\times$ 28.0&20.2 $\times$ 18.3&0.31&0.41&340\\
NGC~6951&39.6 $\times$ 34.9&22.2 $\times$ 18.6&0.35&0.45&338\\
\tableline
NGC~1961&24.6 $\times$ 20.1&18.7 $\times$ 15.0&0.34&0.49&767\\
NGC~3368&44.3 $\times$ 35.6&23.0 $\times$ 19.9&0.28&0.39&399\\
NGC~3718&17.4 $\times$ 15.9&13.9 $\times$ 13.5&0.47&0.84&465\\
NGC~4321&31.3 $\times$ 25.6&18.2 $\times$ 15.9&0.24&0.30&291\\
NGC~4579&43.5 $\times$ 38.6&23.1 $\times$ 21.6&0.33&0.51&374\\
NGC~4736&29.2 $\times$ 28.3&18.9 $\times$ 15.5&0.32&0.28&284\\
NGC~7217&33.4 $\times$ 29.6&18.4 $\times$ 17.0&0.34&0.41&326\\
\tableline
NGC~2782&52.6 $\times$ 48.6&35.1 $\times$ 33.0&0.25&0.37&246\\
NGC~5248&35.5 $\times$ 28.5&19.9 $\times$ 17.5&0.39&0.27&311\\
\tableline
\end{tabular}
\tablecomments{Overview of the observational properties (VLA) beam size and RMS for natural (nat) and robust (rob) weighting. In addition the Full Widths at Zero Intensity (FWZI) of the velocities are listed for all galaxies. The velocity resolution is $\sim$20.8 km s$^{-1}$ (natural) and $\sim$5.2 km s$^{-1}$ (robust). Only NGC~2782 has a velocity resolution of $\sim$10.4 km s$^{-1}$ for robust weighting.}
\label{tab_obs}
\end{table}

\begin{table}
\caption{HI and Optical Morphology}
\begin{tabular}{lcrrrcc}
\tableline\tableline
Name & Morphology& $R_{HI}$ & $R_{opt}$ & $R_{HI}/R_{opt}$ & Disturbed & Companion\\
     & Type & [$\arcsec$] & [$\arcsec$] & & Disk & HI detected\\ 
\tableline 
NGC3147&sp&200&122.1&1.64&&\\
NGC3627&sp&193&267.3&0.72&&\\
NGC4569&c&134&286.5&0.47&&\\
NGC4826&sp/c&578&300&1.93&&\\
NGC5953&c&153&36.9&4.15&x&x\\
NGC6574&c&71&45.3&1.57&&\\
NGC6951&sp&255&97.2&2.62&&\\
\tableline
NGC1961&sp&130&125.1&1.04&x&x\\
NGC3368&r/sp&393&227.4&1.73&x&x\\
NGC3718&warped&460&161&2.86&x&x\\
NGC4321&sp&237&228&1.04&x&\\
NGC4579&r&147&168.6&0.87&&\\
NGC4736&r/sp&427&300&1.42&x&\\
NGC7217&r&121&119.4&1.01&&\\
\tableline
NGC2782&c&120&101.7&1.18&x&\\
NGC5248&sp&334&143.7&2.32&&x\\
\tableline
\end{tabular}
\tablecomments{Overview of the HI morphology compared to the size of the stellar disk. The HI morphology is classified as ring (r), spiral arms (s) and/or bar (b). The values of $R_{opt}$ are taken as d$_{B25}$/2 (Length of the projected major axis at the isophotal level 25 mag/arcsec$^2$ in the B-band ) from LEDA. An equivalent radius for the HI disk $R_{HI}$ is defined at a HI column density of $5.0\times 10^{19}$ cm$^{-2}$. The fraction of the radius of the HI disk ($R_{HI}$) to the radius of the stellar disk ($R_{st}$) is given as $R_{HI}/R_{opt}$. In addition disturbed disks and companions are marked by an "x" (see text for details).}
\label{tab_comp}
\end{table}

\begin{rotate}
\begin{table*}
\caption{HI and Optical Companions}
\begin{scriptsize}
\begin{tabular}{lrcrrrrrr}
\tableline\tableline
Target & log($\sum Q$) &Candidate&Distance&Companion coordinates& $\Delta v_{sys}$& HI flux&HI mass&$\Delta v$ \\
 & &Companions& [kpc] & RA Dec & [km/s] &[Jy km/s]& [$10^9M_{sun}$] & [km/s] \\ 
\tableline 
NGC~3147&-2.8 &UGC 05570 &215&10h20m47.0s +73d17m03s &96& & & \\
NGC~3627&-1.0 &SDSS J111857.87+130500.3 & 38&11h18m57.9s +13d05m00s &136 & & &  \\
        &     &MESSIER 065 &39 &11h18m55.9s +13d05m32s &80 & & & \\
NGC~4569& & - & & & & & & \\
NGC~4826& & - & & & & & & \\
NGC~5953& 0.9 &NGC 5954 & 7 &15h34m35.1s +15d11m54s &-6 & & & \\
        &     &UGC 09902 & 38&15h34m33.2s +15d08m00s &-269 &1.5&0.39 & 110\\
        &     &NGC 5951 &171 &15h33m43.0s +15d00m26s &-185 &10.7&2.75&300\\
        &     &4C +15.49 &204 &15h33m14.4s +15d16m41s & & & &\\
NGC~6574& - & & & & & & \\
NGC~6951& - & & & & & & \\
\tableline
NGC~1961&-1.4 &CGCG 329-011& 119 & 05h43m23.0s +69d25m51s &174 &1.08 &0.74 & 230 \\
        &     &CGCG 329-009 & 141&05h42m36.9s +69d14m12s &-62 & 1.09&0.75 & 210 \\
        &     &             & 163&05h42m46.0s +69d13m3s & -184&0.89 &0.61 & 150 \\
        &     &UGC 03342 & 213&05h44m29.6s +69d17m56s & 40&3.92 &2.69 & 320 \\
        &     &UGC 03344 & 315&05h44m56.6s +69d09m34s & 348&3.93 &2.69 & 150 \\
        &     &UGC 03349 & 422&05h45m38.8s +69d03m38s & 358& & & \\
        &     &CGCG 307-021 &426&05h43m39.7s +68d56m43s &-67& & & \\
NGC~3368&     & - & 37&10h47m43.2s +11d55m50s &78 &0.6&0.01& 40\\
NGC~3718&-1. 2&NGC 3729& 58&11h33m49.3s +53d07m32s & 66& 19.6&1.34 & 250 \\
NGC~4321&-0.8 &NGC 4323 & 26&12h23m01.7s +15d54m20s &262 & & & \\
        &     &IC 0783A & 49&12h22m19.6s +15d44m01s &-364 & & & \\
        &     &VCC 0530 & 56&12h22m07.6s +15d47m57s &-272 & & & \\
        &     &IC 0783  & 92&12h21m38.8s +15d44m42s &-304 & & & \\
NGC~4579&-2.4 &[GKP2005] 106&184&12h38m41.1s +11d58m45s &-483 & & & \\
        &     &VCC 1794 &274&12h39m27.1s +11d46m35s &-291 & & & \\
        &     &VCC 1642 &304&12h35m53.1s +11d40m55s &-189 & & & \\
        &     &NGC 4564 &318&12h36m27.0s +11d26m22s &-377 & & & \\ 
NGC~4736& & - & & & & & & \\
NGC~7217& & - & & & & & & \\
\tableline
NGC~2782&-3.5 &SDSS J091258.41+395540.5 & 183&09h12m58.4s +39d55m41s &-42 & & & \\
NGC~5248&-2.2 &CGCG 073-051 &103 &13h36m43.6s +08d32m48s &-11 & & & \\
        &     &UGC 08575& 117&13h35m45.6s +08d58m09s &10 & 8.36&0.44 & 140\\ 
\tableline
\end{tabular}
\end{scriptsize}
\tablecomments{\scriptsize{Overview of all cataloged companions within a FOV of 60 arcmin and within a velocity difference of $\pm$500 km s$^{-1}$ from the host galaxy (similar to the area probed by the HI data). The position, the projected distance from the host galaxy center and their systemic velocity (NED) relative to the systemic velocities of their host galaxies $\Delta v_{sys}$ are listed. The sum of the tidal strength created by all companions in the field is given as \textbf{$Q$}=log($\sum Q$) (see text for detailed description). A value of $\textbf{Q}<0$ indicates that the tidal forces affecting the primary galaxy are smaller than the internal binding forces, and vice-versa. In case of associated detection in HI gas also the derived HI properties flux, gas mass and velocity width $\Delta v$ are specified.}}
\label{tab_sat}
\end{table*}
\end{rotate}

%

\begin{rotate}
\begin{table}
\begin{scriptsize}
\caption{Kinematic Parameters}
\begin{tabular}{lcrrrrrccc}
\tableline\tableline
Source & RA  & Dec & Offset& $v_{sys}$ & $PA$ & $i$ & $v^{Br}$ &  $R^{Br}$& $n^{Br}$\\
 & (J2000) & (J2000) & [arcsec] & [km/s] & [$deg$] & [$deg$]  & [km/s] & [arcsec] & \\ 
\tableline 
NGC~3147& 10 16 53.66 & +73 24 02.8 & 0.2 $\pm$ 1.0 & 2811.7 $\pm$ 0.6 & 144.6 $\pm$ 1.0 & 34.7 $\pm$ 3.4 & 332$\pm$12 & 61$\pm$22 & 1.03$\pm$0.04 \\
NGC~3627& 11 20 15.59 & +12 59 15.1 & 16.8 $\pm$ 0.7 & 719.7 $\pm$ 6.8 & 173.6 $\pm$ 1.6 & 56.0 $\pm$ 3.2 & 192$\pm$8 & 123$\pm$87 & 1.04$\pm$0.13 \\
NGC~4569& 12 36 49.81 & +13 09 46.0 & 0.3 $\pm$ 1.9 & -228.7 $\pm$ 4.0 & 22.4 $\pm$ 2.9 & 63.7 $\pm$ 3.2 & 167$\pm$64 & 45$\pm$41 & 0.76$\pm$0.07 \\
NGC~4826& 12 56 43.70 & +21 40 54.7 & 2.8 $\pm$ 1.2 & 404.0 $\pm$ 3.3 & 296.2 $\pm$ 6.0 & 48.1 $\pm$ 17.8 & 207$\pm$87 & 53$\pm$122 & 1.06$\pm$0.07 \\
outer disk&             &             &               & 127.4 $\pm$ 2.6 & 64.7 $\pm$ 5.5 & 70$\pm$28 & 1446$\pm$96 & 1.57$\pm$0.21 \\
NGC~5953& 15 34 33.43 & +15 11 53.8 & 22.4 $\pm$ 5.8 & 1906.2 $\pm$ 6.5 & 306.5 $\pm$ 2.8 & 65.2 $\pm$ 2.6 & 139$\pm$640 & 60$\pm$366 & 0.50$\pm$0.20 \\
NGC~6574& 18 11 51.35 & +14 58 55.1 & 1.9 $\pm$ 5.0 & 2285.6 $\pm$ 6.2 & 155.4 $\pm$ 1.9 & 46.3 $\pm$ 3.2 & 196$\pm$30 & 35$\pm$15 & 1.20$\pm$0.15\\
NGC~6951& 20 37 14.26 & +66 06 19.3 & 2.7 $\pm$ 0.5 & 1425.6 $\pm$ 1.3 & 138.7 $\pm$ 1.2 & 41.5 $\pm$ 1.5 & 193$\pm$13.8 & 41$\pm$17 & 0.88$\pm$0.02\\
\tableline
NGC~1961& 05 42 04.35 & +69 22 46.5 & 7.5 $\pm$ 2.4 & 3928.5 $\pm$ 15.8 & 269.1 $\pm$ 5.0 & 42.6 $\pm$ 4.0 & 104$\pm$4 & 274$\pm$1 & 1.95$\pm$0.02 \\
NGC~3368& 10 46 45.53 & +11 49 02.6 & 9.7 $\pm$ 0.5 & 896.9 $\pm$ 2.1 & 169.0 $\pm$ 1.5 & 57.5 $\pm$ 2.1 & 193$\pm$5 & 177$\pm$102 & 1.02$\pm$0.08 \\
NGC~3718& 11 32 34.59 & +53 04 07.9 & 5.2 $\pm$ 0.6 & 986.4 $\pm$ 6.9 & 196.0 $\pm$ 5.7 & 48-60          & 150$\pm$29& 849$\pm$46 & 1.43$\pm$0.11  \\
NGC~4321& 12 22 54.83 & +15 49 20.6 & 1.1 $\pm$ 0.8 & 1577.5 $\pm$ 2.1 & 152.3 $\pm$ 1.4& 31.7 $\pm$ 3.5 & 240$\pm$5 & 176$\pm$117 & 0.99$\pm$0.16 \\
NGC~4579& 12 37 43.64 & +11 49 06.4 & 2.0 $\pm$ 0.5 & 1517.7 $\pm$ 1.0 & 87.2 $\pm$ 1.3 & 44.5 $\pm$ 5.2 & 244$\pm$9 & 83$\pm$50 & 0.97$\pm$0.09 \\
NGC~4736& 12 50 53.00 & +41 07 14.3 & 1.1 $\pm$ 0.1 & 310.4 $\pm$ 0.3 & 303.8 $\pm$ 1.1 & 36.6 $\pm$ 2.4 & 183$\pm$2 & 27$\pm$5 & 0.92$\pm$0.01\\
NGC~7217& 22 07 52.31 & +31 21 33.8 & 1.1 $\pm$ 1.6 & 951.9 $\pm$ 2.5 & 268.0 $\pm$ 1.7 & 33.7 $\pm$ 3.3 & 261$\pm$32 & 50$\pm$64 & 0.94$\pm$0.14\\
\tableline
NGC~2782& 09 14 05.96 & +40 06 53.3 & 13.4 $\pm$ 0.9 & 2559.0 $\pm$ 7.4 & 115 $\pm$ 21.8 & 31.7$^b$ $\pm$ 3.2 & 55$\pm$474 & 456$\pm$448 & 1.69$\pm$4.2\\
NGC~5248& 13 37 32.02 & +08 53 08.1 & 2.0 $\pm$ 0.8 & 1151.1 $\pm$ 1.1 & 114.4 $\pm$ 1.0 & 43.1 $\pm$ 1.6 & 188$\pm$2 & 97$\pm$19 & 0.97$\pm$0.02 \\
\tableline
\end{tabular}
\tablecomments{Overview of the kinematic parameters derived from the observed velocity fields by tilted-ring model fits: The dynamical center of the HI disk and its offset from the optical center, the systemic velocity $v_{sys}$, the position angle $PA$ and the inclination $i$ are listed for all galaxies. A Brandt curve fit has been applied to the rotation velocities where $v^{Br}$ is the maximum rotation velocity, $R^{Br}$ the radius of maximum rotation velocity, and $n$ the Brandt curve index as a parameter of the steepness of the curve. Note that the values for NGC~2782 are likely incorrect, as there is a short tidal tail to the east which is barely resolved at our resolution \citep[see][]{Smi94}.}
\label{tab_kin}
\end{scriptsize}
\end{table}
\end{rotate}

\begin{table}
\caption{Mass Properties}
\begin{tabular}{lrrrrrrr}
\tableline\tableline
Source & Flux  & $M_{HI}$ & $M_{HI}/M_{dyn}$&$v(R_{HI})$ & $R_{HI}$ & $M_{dyn}$& $M_{dyn}^{Brandt}$  \\
 & [Jy km/s] &  [$10^9 M_{sun}$] &  & [km/s] & [kpc] & [$10^{11}M_{sun}$] &  [$10^{11}M_{sun}$]\\ 
\tableline 
NGC3147&22.7&8.96&0.016&249&39.7&5.7&10.1\\
NGC3627&32.8&0.34&0.007&185&6.2&0.5&0.9\\
NGC4569&7.5&0.5&0.003&245&10.9&1.5&1.2\\
NGC4826&45.9&0.18&0.003&149&11.5&0.6&2.5\\
NGC5953&6.9&1.77&0.004&276&24.5&4.3&4.9\\
NGC6574&2.93&0.85&0.019&126&12.0&0.4&1.5\\
NGC6951&34.8&4.77&0.014&221&29.8&3.4&1.7\\
\tableline
NGC1961&68&46.62&0.037&402&34.0&12.8&3.4\\
NGC3368&64.2&0.99&0.010&169&15.4&1.0&2.0\\
NGC3718&115.1&7.85&0.016&234&37.9&4.8&8.6\\
NGC4321&51.1&3.4&0.013&238&19.3&2.5&6.6\\
NGC4579&8.5&0.57&0.004&241&12.0&1.6&3.3\\
NGC4736&87&0.38&0.009&141&8.9&0.4&0.2\\
NGC7217&9.7&0.59&0.004&263&9.4&1.5&2.2\\
\tableline
NGC2782&14.8&4.86&0.046&145&21.7&1.1&1.2\\
NGC5248&81.3&4.32&0.028&166&24.3&1.6&2.0\\
\tableline
\end{tabular}
\tablecomments{Overview of the mass properties, including the measured HI flux, the dynamical mass $M_{dyn}^{Brandt}$ obtained from the Brandt fit and the dynamical mass $M_{dyn}$ derived from the deprojected circular velocity $v(R_{HI})$ at the HI radius $R_{HI}$. The ratio of the HI mass and the dynamical mass $M_{dyn}$ of a galaxy is given by $M_{HI}/M{dyn}$. Note that $M_{dyn}$ for NGC~2782 is likely incorrect, as there is a short tidal tail to the east which is barely resolved at our resolution\citep[see][]{Smi94}.} 
\label{tab_mass}
\end{table}

\begin{table}
\caption{AGN Type Relations}
\begin{tabular}{lccc}
\tableline\tableline
Properties & Seyfert & LINER &  non-AGN  \\
\tableline 
Distance [Mpc]& 22.9 & 20.0 & 26.2 \\
Morph. Type t & 2.8 $\pm$ 0.5 & 2.7 $\pm$ 0.4 & 2.6 \\
Disturbed (\%)& 14 $\pm$ 13 & 71 $\pm$ 17 & 1 \\
HI detected companions (\%)& 14 $\pm$ 13 & 43 $\pm$ 18& 1 \\
$R_{HI}$ [kpc]& 19.2 & 19.6 & 23.0 \\
$R_{B25}$ [kpc] & 12.4 & 14.7 & 14.4 \\
$\frac{R_{HI}}{R_{B25}}$ & 1.9 $\pm$ 0.43 & 1.4 $\pm$ 0.25 & 1.8 \\
HI Mass [$10^9 M_{\sun}$] & 2.5 $\pm$ 1.1 & 8.6 $\pm$ 5.9 & 4.5 \\
$\frac{M_{HI}}{M_{dyn}}$ & 0.009 $\pm$ 0.002 & 0.013 $\pm$ 0.004 & 1.8 \\
Ring (\%)& 0 & 80 $\pm$ 18 & 0 \\
Spiral (\%)& 60 & 60 & 100 \\
\tableline 
\end{tabular}
\tablecomments{Overview of the mean HI properties as a function of AGN type, for Seyferts, LINERs and non-AGN galaxies. For each AGN type the number of galaxies, the average distance, the percentage of galaxies with disturbed disk and HI companions, the mean HI radius $R_{HI}$ at a HI column density of $5.0\times 10^{19}$ cm$^{-2}$, the mean optical radius $R_{B25}$, the ratio $R_{HI}/R_{B25}$, the average HI mass of the disk, the average morphology type $t$, and the percentage of ringed, spiral and concentrated HI structure are listed. In addition, also the corresponding standard deviations  are specified (using Bootstrapping method, see \S \ref{subsec:res_agn}).}
\label{tab_agn}
\end{table}

\clearpage

\end{document}